\documentclass[final]{jfm}

\usepackage{graphicx,tabularx}
\usepackage{newtxtext}
\usepackage{newtxmath}
\usepackage{natbib}
\usepackage{hyperref}
\hypersetup{colorlinks=true,urlcolor= blue,citecolor= black}

\newcommand{\RomanNumeralCaps}[1]
\linenumbers
\usepackage{caption}
\usepackage{subcaption}

\title{Characterization of sea ice kinematics over oceanic eddies}

\author{Minki Kim\aff{1},
  Georgy E. Manucharyan\aff{2}, \and
  Monica M. Wilhelmus\aff{1}\corresp{\email{mmwilhelmus@brown.edu}}}

\affiliation{\aff{1} Center for Fluid Mechanics, School of Engineering, Brown University, Providence, RI 02912, USA
\aff{2} School of Oceanography, University of Washington, Seattle, WA, USA}

\begin{document}
\maketitle


\begin{abstract}
Eddies within the meso/submeso-scale range are prevalent throughout the Arctic Ocean, playing a pivotal role in regulating freshwater budget, heat transfer, and sea ice transport. While observations have suggested a strong connection between the dynamics of sea ice and the underlying turbulent flows, quantifying this relationship remains an ambitious task due to the challenges of acquiring concurrent sea ice and ocean measurements. Recently, an innovative study using a unique algorithm to track sea ice floes showed that ice floes can be used as vorticity-meters of the ocean. Here, we present a numerical and analytical evaluation of this result by estimating the kinematic link between free-drifting ice floes and underlying ocean eddies using idealized vortex models. These analyses are expanded to explore local eddies in quasi-geostrophic turbulence, providing a more realistic representation of eddies in the Arctic Ocean. We find that in both flow fields, the relationship between floe rotation rates and ocean vorticity depends on the relative size of the ice floe to the eddy. As the floe size approaches and exceeds the eddy size, the floe rotation rates depart from half of the ocean vorticity. Finally, the effects of ice floe thickness, atmospheric winds, and floe-floe collisions on floe rotations are investigated. The derived relations and floe statistics set the foundation for leveraging remote sensing observations of floe motions to characterize eddy vorticity at small to moderate scales. This innovative approach opens new possibilities for quantifying Arctic Ocean eddy characteristics, providing valuable inputs for more accurate climate projections. 

\end{abstract}


\section{Introduction} \label{sec:intro}

Mesoscale and submesoscale oceanic eddies, ranging from 10 to 300 km and 0.2 to 20 km, respectively, are widespread throughout the global ocean. These coherent structures are known to contain approximately 80$\%$ of the ocean kinetic energy \citep{Ferrari2009,Klein2019}, influencing global ocean circulation, biogeochemical tracer transport, and energy transfer across various length scales. 
Ocean variability at these scales typically follows the principles of geostrophic turbulence, characterized by an inverse energy cascade that strengthens larger eddies and redistributes kinetic energy and by a forward enstrophy cascade that creates finer filament features in the flow \citep{Boffetta2012,Callies2013,Klein2019}. In ice-covered regions, such as the Arctic Ocean, the characteristics of ocean turbulence are distinct; oceanic eddies contribute to sea ice melting by enhancing vertical mixing and transporting heat from deep ocean layers to the surface \citep{Manucharyan2022a}. In addition, atmosphere-sea ice-ocean interactions lead to momentum and heat exchanges, resulting in increased dissipation of eddy kinetic energy due to sea ice-ocean drag \citep{Liu2024,Muller2024}.

With the continued global warming trend, the Arctic sea ice cover has experienced changes in its characteristics and a rapid decline in its extent \citep{Kwok2009,Rampal2009,Comiso2012}. As a result, Arctic marginal ice zones (MIZ)--the transitional regions between dense pack ice and open ocean--have become more prominent \citep{Strong2013,Rolph2020}, fostering more energetic mesoscale variability in the ocean \citep{Armitage2020,Appen2022}, and intensifying eddy fields \citep{Manucharyan2022b}. However, the variability of MIZ eddies and their relation to sea ice are yet to be fully characterized, partly due to the challenges of acquiring observations in ice-covered regions. On the one hand, while in-situ eddy observations, such as those acquired via Ice-Tethered Profilers \citep{Timmermans2008,Zhao2016} and moorings \citep{Zhao2016,Pnyushkov2018,Cassianides2021}, provide accurate, temporally resolved data, measurements remain sparse and mostly out of MIZ. On the other hand, remote sensing techniques, including satellite altimetry, offer broader spatial coverage and allow for a more comprehensive characterization of the eddy field \citep{Kubryakov2021}. But, processing data in ice-covered areas remains challenging, limiting analyses to seasonally ice-free regions. In addition, satellite altimeters acquire measurements infrequently and with limited spatial coverage at high latitudes, where the Rossby deformation radius is smaller compared to lower latitudes. Notably, the recent launch of the Surface Water and Ocean Topography (SWOT) satellite, with $O$(1 km) spatial resolution, promises more detailed surface velocity information based on altimetry observations \citep{Dibarboure2025}. However, methodologies for deriving surface velocities from sea surface height using SWOT observations are still under development and not yet fully operational, particularly for the MIZ.

Analysis of remote sensing sea ice imagery has been proposed as a unique alternative to standard techniques for the characterization of the turbulent eddy field in MIZ, albeit qualitatively. The first description of subsurface ocean eddies imprinted on ice edges in the Fram Strait MIZ was made using airborne remote sensing imagery \citep{Johannessen1987a}. More recently, studies employing Synthetic Aperture Radar (SAR) images have detected the distinct signature of ocean eddies and filaments on the distribution of sea ice, enabling the quantification of eddy counts, sizes, and positions \citep{Kozlov2019,Cassianides2021,Kozlov2022}. Similarly, under-ice eddy characteristics have been inferred through Lagrangian observations of ice floe rotation rates \citep{Manucharyan2022b}, retrieved from Moderate Resolution Imaging Spectroradiometer (MODIS) optical imagery \citep{Lopez2019}. This study demonstrated that upper ocean eddy vorticity has a tight link to ice floe rotations via ice-ocean torques, as atmospheric winds primarily drive ice floe advection, resulting in persistent daily-scale floe rotations.

From a fundamental standpoint, fluid flows have long been characterized in laboratory settings using micro-size particles as tracers. Inertialess spherical particles, for instance, serve as idealized passive tracers, effectively measuring flow fields via particle image velocimetry \citep{Adrian2011}. However, different particle properties (e.g., inertia, shape, size) result in distinct behaviors. Particle inertia causes a delayed response to changes in the flow fields, leading to misalignment between particle motion and the background flow field \citep{Mortensen2007,Ouellette2008,Zhao2015b,Brandt2022}. In turbulent shear flows, particle inertia can enhance rotational motion, especially near walls \citep{Mortensen2007}. \citet{Ouellette2008} also demonstrated that Lagrangian measurements of inertial particles along their trajectories differ from those of passive tracers in chaotic flows. In addition, inertial particles with different sizes and shapes exhibit unique rotational behaviors induced by local velocity gradients in various types of turbulent flows \citep{Voth2017,Brandt2022}, including homogeneous isotropic turbulence \citep{Bordoloi2017,Allende2023}, turbulent channel flow \citep{Zhao2015b}, and turbulent boundary layers \citep{Tee2024}. These cumulative findings not only advance our fundamental understanding of particle dynamics but also suggest a promising potential for using particles to characterize a wide range of flow fields.

In systems where only a limited number of seeded particles are accessible, Lagrangian approaches using these particles provide an effective way of retrieving flow structure, and examples span cryogenic \citep{Svancara2020} and environmental flows \citep{Dauxois2021}. For instance, in flows involving superfluid $^4$He, where only 100 particles can be detected within a 1 MPixel image, \citet{Outrata2021} employed Lagrangian particles to determine the vorticity of vortex rings.
Similarly, in MIZ, Lagrangian tracking of ice floes presents a valuable opportunity to quantify ocean field characteristics. In general, during winter, when sea ice concentrations are high, eddy activity in the surface layer is weak, with only a limited number of eddies \citep{Zhao2014}. However, in MIZ observed during summer, where many ice floes are present, the eddy field at the surface layer is more active \citep{Kozlov2019,Kubryakov2021,Kozlov2022}, suggesting that Lagrangian observations of ice floes may be useful for characterizing underlying eddies.
Recent efforts have also been aimed at characterizing coherent structures and even flow fields using sparse trajectories of particles \citep{Mowlavi2022,Harms2024} and ice floes \citep{Covington2022}. These cumulative findings in fluid mechanics with particles imply the potential of particle tracking to measure fluid vorticity, suggesting that sea ice could serve as an effective tracer for quantifying ocean flow fields.

In this study, we explore the kinematic relationship between ice floes and underlying ocean eddies, focusing specifically on the role of ice floe size relative to eddy size. We employ ocean eddy models \citep{Arbic2012} and ice floe models \citep{Manucharyan2022c,Montemuro2023} for ice floe-ocean simulations. Ice floe rotation in an idealized ocean vortex is examined, and analytical relations for ice floe rotation are introduced in \S \ref{subsec:results-TG}. The potential applicability of these analyses to more realistic ocean conditions is also discussed in \S \ref{subsec:results-QG}. The effects of key factors, such as ice floe thickness (\S \ref{subsec:diss-thickness}), atmospheric winds (\S \ref{subsec:diss-wind}), floe-floe collisions, and sea ice concentration (\S \ref{subsec:diss-collision}), on the rotational relationship are examined in \S \ref{sec:discussion}.

\section{Methods} \label{sec:methods}

\subsection{Ocean eddy models} \label{subsec:methods-ocn}

Two numerical models were employed to generate idealized ocean flow fields: (a) the Taylor-Green (TG) vortex and (b) a two-layer quasi-geostrophic (QG) model.
We describe the key details of the TG and QG simulations in turn. Simulation parameters for the two models are summarized in table \ref{tab:sim_param}. 

The TG vortex is an idealized two-dimensional flow field frequently employed in the literature due to its exact closed-form solutions for incompressible flows. For example, it has been used as a background flow field in studies of multiphase flows to investigate the motion of particles \citep{Wereley1999,Qiao2015,Jayaram2020}, bubbles \citep{Deng2006}, and droplets \citep{Qiao2014} in idealized flow settings. Here, we examine the motion of ice floes over a TG vortex and derive analytical expressions for their relationship.

We consider a TG vortex cell of size $L_{TG}$ centered at the origin, defined by $\lvert x \rvert,\lvert y \rvert \leq 0.5L_{TG}$. The stream function of the square-shaped TG vortex is given by
\begin{equation} \label{eq:TG_psi}
    \psi_{TG} = -A_{TG}\text{cos} \left(\frac{\pi x}{L_{TG}}\right) \text{cos}\left(\frac{\pi y}{L_{TG}}\right),
\end{equation}
where $A_{TG}$ is the amplitude, and $x$ and $y$ are the spatial coordinates. Maximum vorticity magnitude occurs at the vortex center, gradually decreasing radially outward until reaching zero at the boundary of the vortex cell. Without loss of generality, we consider only the cyclonic case. Both anticyclonic and cyclonic eddies are present in the Arctic Ocean, yet for the results derived here, the anticyclonic and cyclonic cases differ only in sign.


\begin{table*}
\centering
\caption{Parameters and properties for ocean eddy (top) and ice floe (bottom) models.} \label{tab:sim_param}
\begin{tabular}{m{17em} c c c}
\hline
Parameter & Symbol & Definition & Value \\ \hline
Amplitude of stream function (TG)   & $A_{TG}$  & ---   & 1.23 $\times 10^3$ m$^2$/s \\
Size of a vortex cell (TG)          & $L_{TG}$  & ---   & 35 km \\
Domain size (QG)                    & $L_{QG}$  & ---   & 400 km \\
Rossby radius of deformation (QG)   & $L_d$     & ---   & 5.2 km \\
Ratio of layer depths (QG)          & $\delta$  & $H_1/H_2$ & 1 \\
Vertical shear of horizontal currents (QG)      & $\Delta U$ & $U_1 - U_2$ & 0.21 m/s \\
Effective drag length scale (QG)    & $r_1$     & $C_{d,eff}/H_1$ & $2 \times 10^5$ m$^{-1}$ \\
Linear dissipation time scale (QG)  & $r_2$     & ---   & 0.01 days$^{-1}$ \\
\hline
Density (ocean)                     & $\rho_o$  & ---   & 1027 kg/m$^3$ \\
Density (ice floe)                  & $\rho_f$  & ---   & 920 kg/m$^3$ \\
Density (atmosphere)                & $\rho_a$  & ---   & 1.2 kg/m$^3$ \\
Sea ice-ocean drag coefficient      & $C_{d,o}$ & ---   & 5.5 $\times$ 10$^{-3}$ \\
Sea ice-atmosphere drag coefficient & $C_{d,a}$ & ---   & 1.0 $\times$ 10$^{-3}$ \\
Turning angle (ocean)               & $\theta_o$& ---  & 15$^\circ$ \\
Turning angle (atmosphere)          & $\theta_a$& ---  & 0$^\circ$ \\
Radius (ice floe)                   & $R_f$     & ---   & 1--35 km \\
Thickness (ice floe)                & $h_f$     & ---   & 0.1--1.0 m \\
Young's modulus (floe collisions)   & $E_f$     & ---   & 5 $\times$ 10$^7$ Pa \\
Shear modulus (floe collisions)     & $G_f$     & $E_f/2(1+\nu)$ & 1.9 $\times$ 10$^7$ Pa \\
Poisson's ratio (floe collisions)   & $\nu$     & ---   & 0.3 \\
Speed (atmosphere)                  & $\lvert \mathbf{u}_a \rvert$ & --- & 0--12 m/s \\
Coriolis parameter                  & $f$       & ---   & 10$^{-4}$ rad/s$^{-1}$\\
Non-dimensionalized floe inertia    & $H_{f,o}^*$ & $\rho_fh_f/\rho_oC_{d,o}L_o$ & 10$^{-3}$--10$^{-2}$ \\
Rossby number                       & $Ro$      & $U_o/fL_o$ & 0.045 \\
Nansen number                       & $Na$      & $\sqrt{\rho_aC_{d,a}/\rho_oC_{d,o}}$ &0.015 \\
\hline
\end{tabular}
\end{table*}

The two-layer QG model produces more realistic ocean eddies similar in size and shape to those observed in the MIZ.
We employed the QG flow field tuned to observations from the Beaufort Gyre (BG) \citep{Manucharyan2022b}. Full details of the model and the tuning of its parameters can be found in \citet{Arbic2012} and \citet{Manucharyan2022d}. The QG model setup divides the ocean into two vertical layers. The model is forced by an imposed mean flow assumed to be homogeneous in the horizontal direction. The vertically sheared horizontal flows in these two layers induce a baroclinic instability, resulting in the generation of eddies evolving over the horizontal plane of each layer, which is the most common eddy generation mechanism in the global ocean \citep{Tulloch2011}, including the Arctic BG \citep{Hunkins1974,Manucharyan2022d}. 
In both layers, the mean horizontal velocity is imposed in the zonal direction $U$, leading to gradients in the mean potential vorticity, $Q$, which accounts for vorticity with the presence of stratification. These velocity and vorticity gradients result in zonal ($x$) and meridional ($y$) velocity perturbations ($u$ and $v$) as well as potential vorticity perturbation ($q$). The evolution of the perturbation vorticity fields in each layer is described by the QG equations, which account for the Coriolis force and the vertical shear of the velocity. There is no internal friction between layers. The governing equations for each layer are as follows:
\begin{align}
    \frac{\partial q_1}{\partial t} + \left(u_1 + U_1 \right)\frac{\partial q_1}{\partial x} + v_1\frac{\partial q_1}{\partial y} & = -v_1\frac{\partial Q_1}{\partial y} - r_1 \lvert \nabla \times \bf{u_1} \rvert \lvert\bf{u_1}\rvert + \it{s.s.d.}, \label{eq:QG1} \\
    \frac{\partial q_2}{\partial t} + \left(u_2 + U_2 \right)\frac{\partial q_2}{\partial x} + v_2\frac{\partial q_2}{\partial y} & = -v_2\frac{\partial Q_2}{\partial y} - r_2\nabla^2 \psi_2 + \it{s.s.d.}, \label{eq:QG2}
\end{align}
where subscripts $1$ and $2$ denote the top and bottom layers, respectively, $t$ is the time, $\bf{u}$ is the vector form of the perturbed velocity, $r_1 = C_{d,eff}/H_1$ stands for the effective drag length scale for the top layer, $C_{d,eff}$ is the effective sea ice-ocean drag coefficient, $H$ is the layer depth, $r_2$ is the linear dissipation time scale for the bottom layer, $\psi$ is the perturbation stream function, and $s.s.d.$ is small-scale dissipation using an exponential cutoff filter. In this set of equations, uniform dissipation caused by the quadratic sea ice-ocean drag is incorporated in the entire horizontal domain of the top layer, while Ekman-type friction is included in the bottom layer. Small-scale dissipation is also considered to prevent a forward-enstrophy cascade towards small scales. The imposed mean potential vorticity gradients are given by:
\begin{equation}
    \frac{\partial Q_1}{\partial y} = \frac{U_1-U_2}{(1+\delta)L_d^2}, \qquad 
    \frac{\partial Q_2}{\partial y} = \frac{\delta \left(U_2-U_1\right)}{(1+\delta)L_d^2},
\end{equation}
where $\delta = H_1/H_2$ is the ratio of layer depths, $L_d$ is the Rossby radius of deformation, and $\Delta U=U_1-U_2$ is the vertical shear of horizontal currents. The perturbation potential vorticities in the two layers are given by:
\begin{equation}
    q_1 = \nabla^2 \psi_1 +\frac{\psi_2-\psi_1}{(1+\delta)L_d^2}, \qquad 
    q_2 = \nabla^2 \psi_2 +\frac{\delta(\psi_1-\psi_2)}{(1+\delta)L_d^2}.
\end{equation}
The model uses an $f$-plane approximation since the Coriolis parameter is approximately constant in the Arctic Ocean \citep{Timmermans2020}.

The simulation domain spans 400 km $\times$ 400 km and is set up with doubly-periodic boundary conditions. Time integration was performed using the Adams-Bashforth two-step method in Fourier space with 256 modes, producing a converged energy spectrum previously validated by \citet{Manucharyan2022b}. The model was initialized with randomly generated $q_1$ and $q_2$ in Fourier space. The simulation was conducted until the flow field reached an equilibrated state. At that point the energy production from the mean flow was balanced by the energy dissipation from the top and bottom layers due to the sea ice-ocean drag and the Ekman drag, respectively. This equilibration state is typically achieved after approximately one simulation year. The model tuning parameters, $L_d$, $\delta$, and $\Delta U$, were adopted from \citet{Manucharyan2022b}. As a result, the simulated eddy fields closely matched the statistics of the BG MIZ between 2003 and 2020. It is important to note that the sizes of the produced eddies are within the observed range of the eddy sizes (on the order of 10 km) in the MIZ \citep{Johannessen1987a,Kozlov2019,Kozlov2022}. In addition, the estimated eddy kinetic energy derived from the simulated ice floes is comparable to the estimated eddy kinetic energy from in situ measurements via moorings located in the area of the satellite observations \citep{Manucharyan2022b}.

Arctic MIZ eddies typically persist for a period of $O(10)$ days \citep{Johannessen1987a,Kozlov2020,Cassianides2021,Kozlov2022}; hence, we consider stationary ocean fields during the 30-day simulation period. 
Note that the simulation incorporates the effects of sea ice-ocean drag as an effective, continuous, stationary drag force over the top layer, influencing the energetics of the QG eddy field. The quadratic drag law is used due to the turbulent nature of the flow field, consistent with studies on the ice-ocean boundary layer \citep{Mcphee2012,Cole2014}. The effective sea ice-ocean drag coefficient represents the overall impact of drag forces on the ocean field, implying the product of sea ice concentration and the actual ice-ocean drag coefficient \citep{Manucharyan2022d}. This constant coefficient neglects small-scale floe dynamics and seasonal variations. 
For the given Rossby radius, changes in the effective drag coefficient of less than $25\%$ compared to the tuned value have negligible effects on the slope of the eddy energy spectrum. This implies that such variations in ice floe surface properties may have a limited impact on energy transport across scales.

While the tuning parameters in the QG model can produce consistent eddy energetics, local eddy sizes and velocities may vary, even for simulation runs set up with identical tuning parameters. Therefore, instead of covering a wide range of eddy sizes, we set the parameters in the TG vortex model according to the relative length scales of eddies and ice floes (table \ref{tab:sim_param}). The majority of the observed ice floe sizes range from 1 to 35 km \citep{Manucharyan2022b}. $L_{TG}$ was set to broadly cover the floe-eddy size ratios from 0.05 to 2. Ratios higher than 2 were excluded, as these larger floes tend to filter out most of the eddy information and have a limited reflection of the local eddies underneath them. Concurrently, $A_{TG}$ in equation \ref{eq:TG_psi}, was chosen to represent the maximum velocity of local QG eddies corresponding to $L_{TG}$. These length and velocity scales fall within the ranges of observed eddy sizes (1--40 km) \citep{Kozlov2019} and flow speeds (0--0.5 m/s) \citep{Kozlov2022} in the BG MIZ.

Finally, local eddies (or coherent vortices) in ocean flow fields are identified in this study by using the Lagrangian-averaged vorticity deviation (LAVD)-based approach proposed by \cite{Haller2016}. We note that other identification schemes exist, such as the Okubo-Weiss parameter \citep{Okubo1970,Isern2004}, Lagrangian trajectory methods \citep{Haller2005,Dong2011}, and others based on the values of sea surface heights, and vorticity \citep{Chelton2011,Mason2014}.
We opted for the LAVD-based eddy detection method given that it is both time- and rotation-invariant and has demonstrated good performance in detecting coherent structures in altimeter-derived velocity fields of the global ocean (AVISO: Archiving, Validation and Interpretation of Satellite Oceanographic data) \citep{Abernathey2018,Liu2023}. More details are provided in \S\ref{subsec:results-QG}. 


\subsection{Sea ice model} \label{subsec:methods-sea ice}
We simulate the motion of circular ice floes using the SubZero discrete element sea ice model \citep{Manucharyan2022c}. This model is designed to study the behavior of ice floes under mechanical forcing. We parameterize the sea ice-ocean and sea ice-atmosphere stresses through a quadratic drag law \citep{Lepparanta2011}:
\begin{equation} \label{eq:stress_quad}
    \boldsymbol{\tau}_o = \rho_oC_{d,o} \lvert \bf{u_{\it{o}}} - \bf{u_{\it{i}}} \rvert \it e^{i\theta_{o}} \left(\bf{u_{\it{o}}} - \bf{u_{\it{i}}} \right) \it, \quad
    \boldsymbol{\tau}_{a} = \rho_aC_{d,a} \lvert \bf{u_{\it{a}}} - \bf{u_{\it{i}}} \rvert \it e^{i\theta_{a}} \left(\bf{u_{\it{a}}} - \bf{u_{\it{i}}} \right),  
\end{equation}
where $\boldsymbol{\tau}$ is the shear stress, $\rho$ is the density, $\theta$ is the turning angle, and the subscript $o$, $a$, and $i$ correspond to the ocean, the atmosphere, and the ice floe, respectively. Since $\bf{u_{\it{a}}} \gg \bf{u_{\it{i}}}$, $(\bf{u_{\it{a}}} - \bf{u_{\it{i}}}) \approx \bf{u_{\it{a}}}$ was used. The ocean turning angle is the angle between the geostrophic current and the surface shear stress, as the Coriolis effect causes the flow to turn within the boundary layers. Here, $\mathbf{u}_{\textit{o}}$ represents the velocity of the geostrophic current beneath the Ekman layer ($\sim$20 m \citep{Yang2006,Ma2017}), with a direction rotated relative to the ocean surface velocity by a turning angle of $\theta_{o}$. The geostrophic velocity generally has larger magnitudes than the surface ocean velocity in the BG due to sea ice-ocean drag. The wind turning angle is set to zero, as the surface wind is considered.

The velocity of ice floes comprises translational and rotational components, $\bf{u_{\it{i}}} = \bf{u_{\it{f}}} + \rm \Omega_{\it{f}}\hat{\bf{k}} \times \bf{r'}$, where $\bf{u}_{\it{i}}$ is the sea ice velocity over the ice floe area, $\bf{u}_{\it{f}}$ is the translational velocity of the center-of-mass, $\Omega_f$ is the rotation rate of the ice floe, and $\bf{r}' = \bf{r}-\bf{r_{\it{C}}}$ is the position vector on the ice floe with respect to its center-of-mass, $C$. The translational velocity and rotation rate of ice floes evolve according to the linear and angular momentum conservation equations:
\begin{align}
    M_f\left(\frac{\mathrm{d}\mathbf{u}_f}{\mathrm{d}t} + f\hat{\mathbf{k}} \times \mathbf{u}_f \right) = & \iint_A \left( \boldsymbol{\tau}_{o} + \boldsymbol{\tau}_{a} - \rho_f h_f g\nabla\eta\right)\mathrm{d} A + \sum_{j,k} \mathbf{F}_{j,k}, \label{eq:sea-ice-eqns_force} \\
    I_f\frac{\mathrm{d}\Omega_f}{\mathrm{d}t} = & \iint_A \mathbf{r}' \times \left( \boldsymbol{\tau}_{o} + \boldsymbol{\tau}_{a} - \rho_fh_fg\nabla\eta\right)dA + \sum_{j,k} \mathbf{r}_{j,k}' \times \mathbf{F}_{j,k}, \label{eq:sea-ice-eqns_torque}
\end{align}
where $M_f=\rho_fA_fh_f$ is the floe mass, $\rho_f$ is the floe density, $A_f = \pi R_f^2$ is the floe area, $h_f$ is the floe thickness, $R_f$ is the floe radius, $t$ is time, $g$ is the gravitational acceleration, $\eta$ is the sea surface height anomaly associated with ocean currents, $A$ is the surface area covered by an ice floe, $\bf{F}_{\it{j,k}}$ is the interaction forces of the $k$th contact point with the $j$th ice floe due to collisions, $I_f = M_fR_f^2/2$ is the moment of inertia for a floe with an axis passing through the center-of-mass, and $\bf{r}_{\it{j,k}}' = \bf{r}_{\it{j,k}} - \bf{r_{\it{C}}}$ is the position vector on the $k$th contact point with the $j$th ice floe with respect to its center-of-mass. The left-hand side of \eqref{eq:sea-ice-eqns_force} represents the rate of change of ice floe momentum and the Coriolis force acting on ice floes, while the integration terms on the right-hand side consist of the sea ice-ocean stress, the sea ice-atmosphere stress, and the pressure gradient due to the sea surface tilt. Equation \eqref{eq:sea-ice-eqns_torque} consists of the corresponding torque terms. Floe-floe collisions are neglected for ice floes subjected only to oceanic and wind forcings, resulting in $\mathbf{F}_{\it{j,k}} = 0$. 
While these floes can be more observable in regions with low sea ice concentrations, they effectively capture the key connections between ice floes and underlying eddies. 
In addition, it establishes a baseline for ice floe motions in regions with high concentrations. 
Free-drifting ice floes with no collisions are examined in most sections, while the effects of floe-floe collisions are discussed in \S\ref{subsec:diss-collision}. 

Using the length $L_o$, velocity $U_o$, and time $T_o$ scales of the ocean field, \eqref{eq:sea-ice-eqns_force} and \eqref{eq:sea-ice-eqns_torque} can be rewritten in non-dimensionalized form:
\begin{align} 
\begin{split} \label{eq:sea-ice-eqns_force_non}
    \frac{\mathrm{d}\bf{u_{\it{f}}^*}}{\mathrm{d}t^*} = & \frac{1}{H_{f,o}^*} \iint_{A^*} \lvert \mathbf{u}_{o}^* - \mathbf{u}_{i}^* \rvert e^{i\theta_{o}} \left(\mathbf{u}_o^* - \bf{u_{\it{i}}^*} \right) \mathrm{d}A^* \\  & + \frac{Na^2}{H_{f,o}^*}\iint_{A^*} \lvert \bf{u_{\it{a}}^*} - \bf{u_{\it{i}}^*} \rvert \it \left(\bf{u_{\it{a}}^*} - \bf{u_{\it{i}}^*} \right) \mathrm{d}A^* + \frac{\rm{1}}{Ro} \iint_{A^*} \hat{\bf{k}} \times (\bf{u_{\it{o}}^*} - \bf{u_{\it{f}}^*}) \it \mathrm{d}A^*, \\
\end{split} \\
\begin{split} \label{eq:sea-ice-eqns_torque_non}
    &\frac{\mathrm{d}\Omega_{f}^*}{\mathrm{d}t^*} = \frac{2}{R_f^{*2}} \Bigg[ \frac{1}{H_{f,o}^*} \iint_{A^*} \mathbf{r}'^* \times \left[\lvert \mathbf{u}_o^* - \mathbf{u}_i^* \rvert \it e^{i\theta_{o}} \left(\mathbf{u}_o^* - \mathbf{u}_i^* \right)\right] \mathrm{d}A^* \\  & + \frac{Na^2}{H_{f,o}^*}\iint_{A^*} \mathbf{r}'^* \times \left[\lvert \mathbf{u}_a^* - \mathbf{u}_i^* \rvert \it \left(\mathbf{u}_a^* - \mathbf{u}_i^* \right)\right] \mathrm{d}A^* + \frac{\mathrm{1}}{Ro}\iint_{A^*} \mathbf{r}'^* \times \left(\hat{\mathbf{k}} \times \mathbf{u}_o^* \right) \mathrm{d}A^* \Bigg],
\end{split}
\end{align}
where $H_{f,o}^* = \rho_fh_f/\rho_oC_{d,o}L_o$ is the non-dimensionalized floe inertia characterizing ice floe inertia relative to surface ocean drag, $Na=\sqrt{\rho_aC_{d,a}/\rho_oC_{d,o}}$ denotes the Nansen number, $Ro = U_o/fL_o$ denotes the Rossby number, $A^* = A/A_f$ is the non-dimensionalized floe area, and the superscript $*$ indicates non-dimensionalized quantities. Note that the sea surface height tilt terms in \eqref{eq:sea-ice-eqns_force} and \eqref{eq:sea-ice-eqns_torque} are rewritten using $-g\nabla \eta = f\hat{\bf{k}} \times \bf{u_{\it{o}}}$, indicating the geostrophic balance between the pressure gradient force due to the sea surface tilt and the Coriolis force. Interaction forces due to floe-floe collisions are neglected in the non-dimensionalized forms. 

Simulations with the SubZero model employ an Adams-Bashforth two-step method for time integration and a Monte-Carlo scheme \citep{Caflisch1998} at each time step for spatial integration of forces and torques acting on individual ice floes. The velocity and rotation rate of ice floes are initialized with the averaged values of ocean eddies beneath them. In this study, we excluded the first five days of simulation to eliminate any influence of initial conditions on the dynamics. An extensive description of the Subzero model can be found in \cite{Manucharyan2022c,Montemuro2023}. The codes associated with the model are available at \href{https://github.com/SeaIce-Math/SubZero}{https://github.com/SeaIce-Math/SubZero}.

The physical properties of the ocean, sea ice, and the atmosphere used in the simulations are chosen based on geostrophic drag coefficients and turning angles specific to conditions in the BG MIZ (table \ref{tab:sim_param}, from \cite{Lepparanta2011} and \cite{Brenner2021}). We considered ice floes with sizes ranging from 1 to 35 km, covering most of the observations acquired in the BG MIZ from 2003 to 2000 \citep{Manucharyan2022b}. We removed the influence of shape variations on the rotational relationship between ice floes and the ocean by reducing observed floe geometries to circular shapes. This agrees with studies by \cite{Gupta2022} and \cite{Gupta2024}. For non-dimensionalization, the size and the velocity amplitude of a TG vortex are taken as the reference length and velocity scales, respectively.

For the initial part of this study, we considered ice floes with a constant thickness of $h_f = 0.5$ m, in agreement with observed values in the BG MIZ \citep{Krishfield2014,Timmermans2020,Manucharyan2022b}. We also neglected atmospheric stresses. The effects of varying ice floe thickness and wind speeds on the rotational relationship between ice floes and underlying ocean eddies are presented in \S\ref{subsec:diss-thickness} and \S\ref{subsec:diss-wind}. Lastly, the passive tracer, devoid of inertia, perfectly follows the fluid, mirroring the local velocity and rotation of fluid flows. The passive tracer scenario serves as an idealized baseline case in our analysis.

\section{Results} \label{sec:results}

This section investigates the rotational relationship between ice floes and underlying ocean eddies for different floe-eddy size ratios in the TG vortex and two-layer QG flow fields. Two types of ocean quantities were explored for the analyses: (i) area-averaged ocean quantities, calculated by averaging ocean quantities over the ice floe area, representing localized ocean information in regions with ice floe coverage, and (ii) center-of-mass ocean quantities, obtained through the interpolation of quantities at the center-of-mass of the ice floe, offering pointwise ocean information. These two types of quantities are complementary; area-averaged quantities can be used to create spatial vorticity maps when sea ice concentration is high, while quantities derived from information at the center-of-mass can be leveraged to estimate ocean vorticity at lower concentrations, mainly when ice floes undergo a closed-loop trajectory over a larger area. 

\subsection{Single ice floe dynamics} \label{subsec:results-single ice floe}

\begin{figure}
  \centerline{\includegraphics[width=\textwidth]{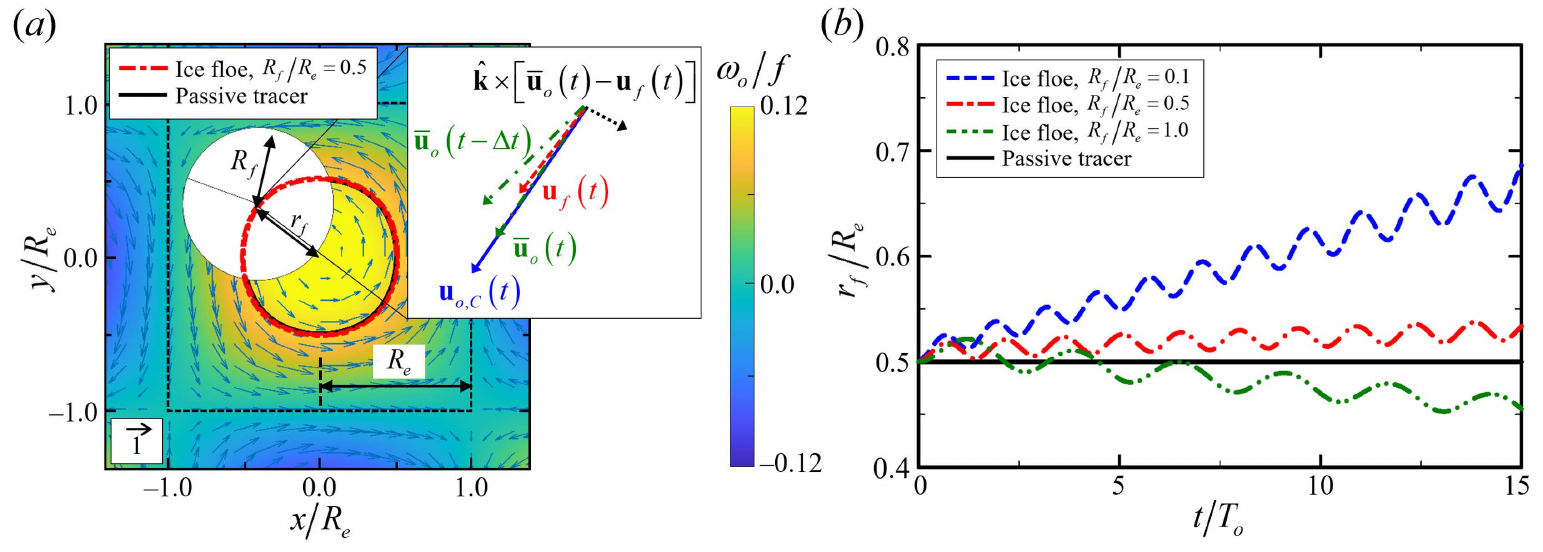}}
  \caption{Ice floes drifting over a TG vortex field. ($a$) The trajectory of a circular ice floe released at the radial position, $r_f$, set to be half of the TG vortex size, $R_e$ ($R_f/R_e = 0.5$). The colors map the magnitude of the fluid vorticity normalized by the Coriolis parameter, while the arrows indicate the direction of the fluid velocity (with the arrow size set to scale according to the velocity magnitude) at a given location. The inset schematic shows the orientation of relevant vectors: the ocean velocity averaged over the floe area, $\overline{\bf{u}}_{\it{o}}$, the ocean velocity at the center-of-mass of the floe, $\bf{u}_{\it{o,C}}$, the ice floe velocity at its center-of-mass, $\bf{u}_{\it{f}}$, and the force direction resulting from combining the Coriolis force and the pressure gradient force due to the sea surface tilt, $\hat{k}\times(\overline{\bf{u}}_{\it{o}} - \bf{u}_{\it{f}})$. ($b$) Radial positions of ice floes with floe eddy size ratios of $R_f/R_e=$ 0.1 (blue dashed lines), 0.5 (red dot-dashed lines), and 1.0 (green double dot-dashed lines), normalized by the size of the eddy. The ice floe cases are compared to the passive tracer case (black solid lines). Supplementary movies are available.} \label{Fig:TG_traj}
\end{figure}

The motion of individual ice floes with different sizes was analyzed in a TG vortex field and compared to the passive tracer case (figure \ref{Fig:TG_traj}). 
Three floe-eddy size ratios were considered, $R_f/R_e=0.1, 0.5,$ and $1.0$.
In all cases, the center-of-mass of the ice floes was initially positioned at $r_f = 0.5R_e$, where $r_f$ denotes the radial position of the floe center-of-mass, to capture the influence of ocean fields both inside and outside the vortex cell.

The trajectories of free-drifting ice floes are shaped by ocean and atmospheric forcing, the effects of floe inertia, the Coriolis force, and the pressure gradient force due to the sea surface tilt. Under low to negligible wind speeds and low sea ice concentration, ice floe motion is expected to be predominantly driven by oceanic forcing. However, it has been hypothesized that ice floe inertia plays an important role in setting the direction in which floes translate by delaying their response to changes in the underlying ocean flow field. From the cases tested, the ice floe with a size ratio of $R_f/R_e=0.5$ exhibited the best performance at resembling a passive tracer and forming a closed-loop trajectory over the TG vortex cell (figure \ref{Fig:TG_traj}$a$). While inertia does result in an outward tilt of the ice floe velocity $\mathbf{u}_f(t)$ relative to the ocean velocity averaged over the floe area at $t$, $\overline{\mathbf{u}}_o(t)$ (a subset of figure \ref{Fig:TG_traj}$a$), the resulting force from combining the Coriolis force and the pressure gradient force due to the sea surface tilt, $\hat{k}\times(\overline{\bf{u}}_{\it{o}} - \bf{u}_{\it{f}})$, reduces this spiraling effect, effectively driving the ice floe to form a closed loop (figure \ref{Fig:TG_traj}$a$). Finally, ice floes were observed to undergo alternating periods of acceleration and deceleration relative to the equilibrium state in which all forces balance instantaneously to zero. As a result, the radial component of floe position oscillates (figure \ref{Fig:TG_traj}$b$).


Changes in floe size with respect to vortex size result in a deviation of floe motion with respect to the eddying motion underneath. For example, smaller ice floes, for which $R_f/R_e=0.1$, followed an outwardly spiral trajectory. In this case, the ice floe velocity and the averaged ocean velocity vectors were better aligned, reducing the inward effect from the Coriolis and pressure gradient forces compared to floes with larger size ratios. This is similar to the behavior of millimeter-sized particles in a Taylor vortex \citep{Wereley1999,Deng2006,Qiao2015} and a Rankine vortex \citep{Varaksin2022}, in which the role of inertia is linked to their resulting spiral trajectories relative to the flow direction. As the floe-eddy size ratio increases to $R_f/R_e=1.0$, discrepancies between ice floe and averaged ocean velocities become more pronounced due to filtration of measured ocean velocities over larger floe areas. This results in larger forces directed toward the vortex center compared to smaller floes, leading to an inward spiraling motion.

\begin{figure}
  \centerline{\includegraphics[width=\textwidth]{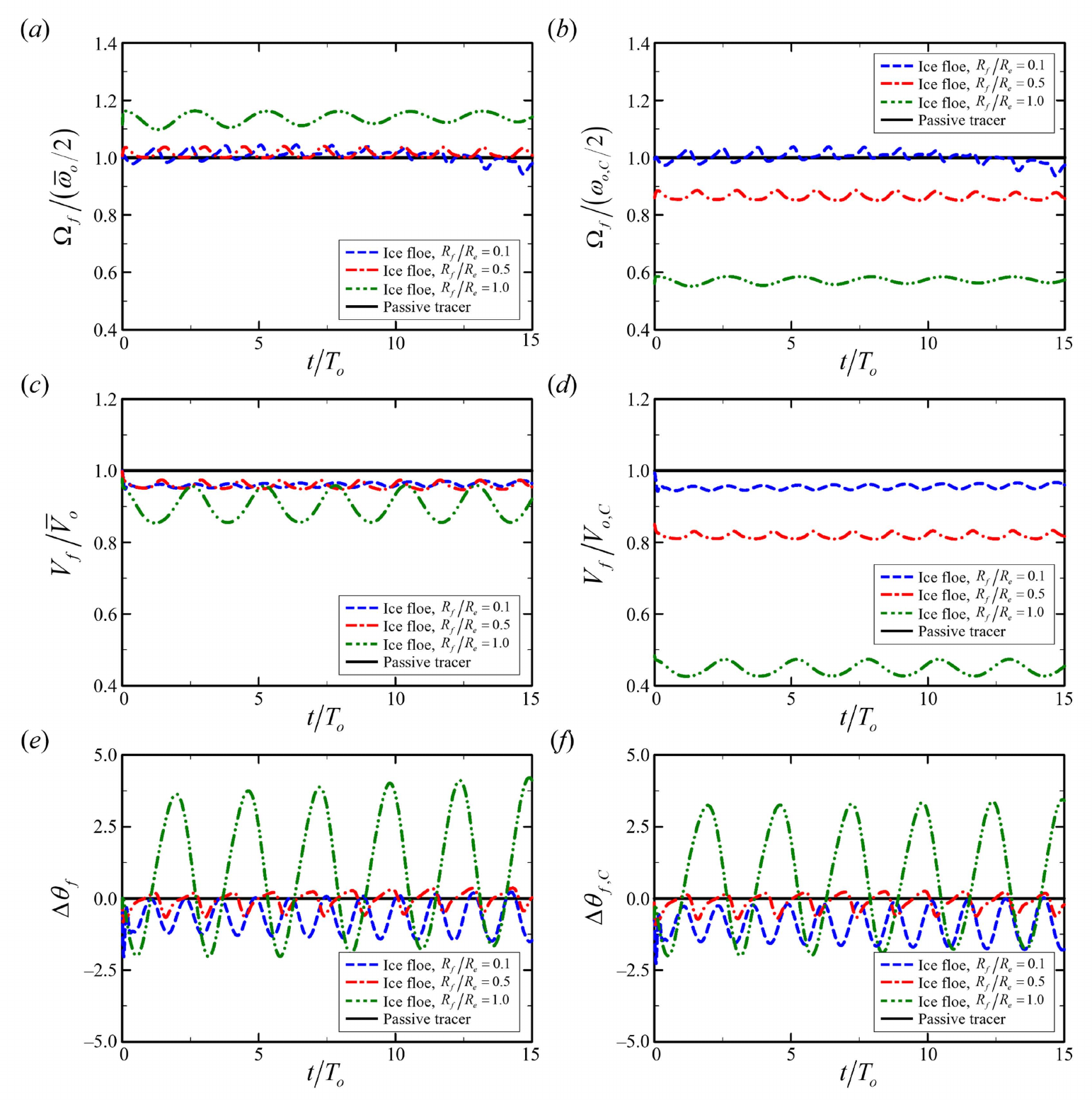}}
  \caption{Motions of ice floes with different sizes in a TG vortex field. ($a,b$) Normalized rotation rates, $\Omega_f$, ($c,d$) speeds, $V_f$, and ($e,f$) angle deviations of ice floes for floe-eddy size ratios of $R_f/R_e=$ 0.1 (blue dashed lines), 0.5 (red dot-dashed lines), and 1.0 (green double dot-dashed lines) are compared to the passive tracer case (black solid lines). Floe rotation rates are normalized by the ($a$) ocean vorticity averaged over the floe area, $\overline{\omega}_o$, and ($b$) ocean vorticity at the floe center-of-mass, $\omega_{o,C}$. Floe speeds are normalized by ($c$) ocean speed averaged over the floe area, $\overline{V}_o$, and ($d$) ocean speed at the floe center-of-mass, $V_{o,C}$. Angle deviations are calculated by subtracting the angle of the ($e$) averaged ocean velocity, $\Delta \theta_f$, and the ($f$) center-of-mass ocean velocity, $\Delta \theta_{f,C}$, from that of the ice floe velocity. The ice floes are initially released at $r_f/R_e = 0.5$.} \label{Fig:T_F_time_evol}
\end{figure}

The kinematics of ice floes with different sizes were examined along their trajectories (figure \ref{Fig:T_F_time_evol}). We present results using two normalizations: (i) considering ocean quantities averaged over the floe area, including averaged ocean vorticity, $\overline{\omega}_o$, and averaged ocean speed, $\overline{V}_o$; and (ii) ocean quantities at the floe center-of-mass, such as ocean vorticity, $\omega_{o,C}$, and ocean speed, $V_{o,C}$. Note that the floe rotation rates are normalized by half of the ocean vorticity. Angle deviations are calculated by subtracting the angle of the ocean velocity from that of the ice floe velocity, $\Delta \theta_f = \theta_f - \overline{\theta}_o$ and $\Delta \theta_{f,C} = \theta_f - \theta_{o,C}$, where $\theta_f$ denotes the angle of the ice floe velocity, $\overline{\theta}_o$ denotes the angle of the ocean velocity averaged over the ice floe area, and $\theta_{o,C}$ denotes the angle of the ocean velocity at the center of mass of the ice floe. All angles are calculated using the direction of each velocity vector with respect to the positive $x$-axis, measured in the counter-clockwise direction.

Normalized ice floe rotation rates, velocities, and orientations oscillated over time, periodically crossing the equilibrium values. Small ice floes with $R_f/R_e = 0.1$ show higher sensitivity to local ocean quantities, leading to normalized rotation rates and speeds close to unity and minimal angle deviations (figure \ref{Fig:T_F_time_evol}). Ocean quantities averaged over the floe area and at the floe center-of-mass exhibit similar values, resulting in comparable normalized rotation rates (figures \ref{Fig:T_F_time_evol}$a$ and \ref{Fig:T_F_time_evol}$b$) and speeds (figures \ref{Fig:T_F_time_evol}$c$ and \ref{Fig:T_F_time_evol}$d$). As ice floes move farther from the vortex center along their outwardly spiral trajectories, the normalized rotation rates begin to decrease at later times. The angle deviations have negative values for most of the evolution time due to the outwardly spiral shape of the trajectory (figures \ref{Fig:T_F_time_evol}$e$ and \ref{Fig:T_F_time_evol}$f$). Contrasting the ice floe case, the passive tracer perfectly mirrors the vorticity and velocity of the underlying flow field with zero angle deviations.

As the floe-eddy size ratios increase to $R_f/R_e = 1.0$, the normalized ice floe rotation rates and speeds exhibit larger deviations from unity (figures \ref{Fig:T_F_time_evol}$a$--$d$). 
With larger floe-eddy size ratios, the floe rotation rates normalized by the averaged ocean vorticity increase because the floe samples low-vorticity regions near cell boundaries, reducing the averaged ocean vorticity (figure \ref{Fig:T_F_time_evol}$a$). 
Conversely, the floe speeds normalized by the averaged ocean speed decrease due to high-velocity regions near the cell boundaries covered by the floe area, resulting in higher averaged ocean speed (figure \ref{Fig:T_F_time_evol}$c$). 
In the context of ocean quantities at the floe center-of-mass, the normalized rotation rates and speeds of ice floes decrease significantly for larger floe-eddy size ratios. 
This occurs because the ocean vorticity and velocity at the floe center-of-mass are considerably higher compared to the values averaged over the floe area (figures \ref{Fig:T_F_time_evol}$b$ and \ref{Fig:T_F_time_evol}$d$). 
For both types of ocean quantities, the angle deviation becomes more pronounced as the floe-eddy size ratios increase (figures \ref{Fig:T_F_time_evol}$e$ and \ref{Fig:T_F_time_evol}$f$). 
With larger ratios, the resultant forces from a combination of the Coriolis force and the pressure gradient force due to the sea surface tilt become larger, resulting in a shift of floe trajectories from an outwardly spiral shape to an inwardly spiral shape as the ratios increase. 
As a result, the angle deviations at $R_f/R_e=0.5$ are smaller than in the other two cases since the floe trajectories form nearly closed loops. 
Floes with $R_f/R_e=1.0$ show positive angle deviations since they have inwardly spiral trajectories.

The motion of a single floe provides insight into the instantaneous response of ice floes to an underlying ocean eddy flow field. Our results demonstrate the importance of floe-eddy size ratios modulating the distribution of forces due to drag, Coriolis, and the pressure gradient on ice floes as they drift over ocean eddy fields.

\subsection{Statistics of ice floe kinematics in an idealized vortex} \label{subsec:results-TG}

Based on the statistics of ice floes in a TG vortex field, we derived analytical expressions describing the kinematic link between ice floe rotation and the ocean vorticity underneath. Appendix \ref{sec:app_eddy_identification} outlines the criteria for identifying trapped ice floes. Given the time duration of the simulation and the requirements for floe selection, ice floes with floe-eddy size ratios ranging from 0.05 to 1.4 were considered for further analysis. 

\begin{figure}
  \centerline{\includegraphics[width=\textwidth]{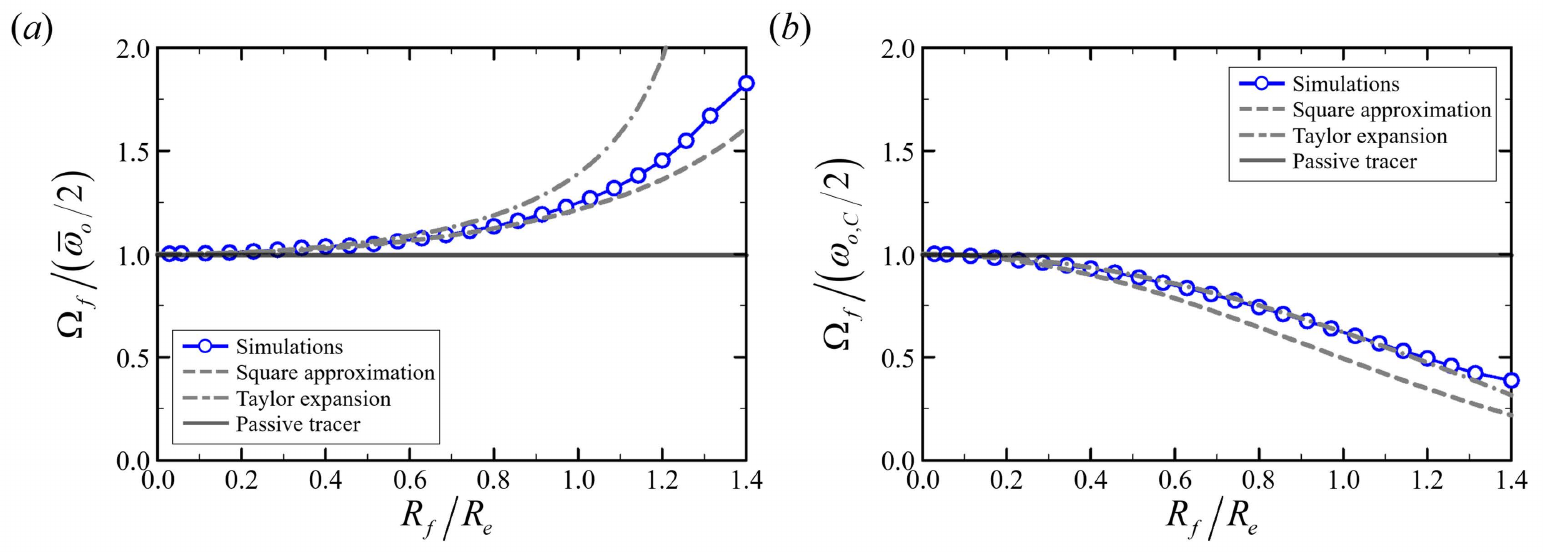}}
  \caption{Rotational motion of centered ice floes in a TG vortex field. Ice floe rotation rates, $\Omega_f$, normalized by the ($a$) ocean vorticity averaged over the floe area, $\overline{\omega}_o$, and the ($b$) the ocean vorticity at the floe center-of-mass, $\omega_{o,C}$, for different floe-eddy size ratios, $R_f/R_e$. The simulation results are compared to analytical relations using square-shape approximation (dashed lines) and Taylor series expansion (dashed-dotted lines), and the passive tracer case (solid lines).} \label{Fig:TG_cen_rot}
\end{figure}

We examined ice floes positioned at the center of a TG vortex representing an idealized scenario (figure \ref{Fig:TG_cen_rot}).
In this case, the ice floes are initially released at the center of the vortex, exhibiting a distinct trend in their rotation driven solely by spatial variations in ocean vorticity within the vortex core. The ice floe rotation rate normalized by the ocean vorticity averaged over the floe area shows a monotonic growth with increasing floe-eddy size ratios (figure \ref{Fig:TG_cen_rot}$a$).
This observed rise in rotation rates can be attributed to the gradual decline in local ocean vorticity from the center of a vortex cell toward its periphery.
As floe-eddy size ratios increase, the averaged ocean vorticity decreases,increasing normalized rotation rates. In contrast, the rotation rate normalized by the ocean vorticity at the floe center-of-mass decreases for larger floe-eddy size ratios (\ref{Fig:TG_cen_rot}$b$). This behavior can be attributed to the fact that large ice floes rotate at slower rates while the ocean vorticity is maximized at the center-of-mass of the rotating ice floes.

Building upon the physical interpretation of ice floe rotation, we established analytical relations for the normalized rotation rates by balancing the torques acting on the ice floes. 
In the present analysis, the quadratic drag terms in the angular momentum equation \eqref{eq:sea-ice-eqns_torque_non} are substituted with linear drag, thereby enabling the derivation of explicit relations describing the dependence of the normalized rotation rates on floe-eddy size ratios.
We investigated the effect of this substituted parameterization on the rotational relationship using the Rankine vortex, a simplified case for which solutions for both drag parameterizations can be derived.
The analytical relation for the normalized rotation rate of ice floes shows minimal sensitivity to the choice of drag parametrization, as detailed in Appendix \ref{sec:app_RK}, such that a linear drag law was used in deriving an analytical relation.

By introducing a so-called size parameter, $\gamma = R_f/R_e$,  we derived an analytical relationship for the normalized rotation rate of an ice floe located at the center of the TG vortex. Near the center of the TG vortex ($2\pi x/L$, $2\pi y/L \ll 1$), the ocean surface velocity can be approximated using a Taylor series expansion. In the absence of wind, ice floe rotation is dominated by ice-ocean stress for which  the induced equilibrium of torques leads to the relation: 
\begin{equation}
    \iint_{A^*} \bf{r}_{\it{f}}'^* \times (\bf{u}_{\it{f}}^* \it - \bf{u}_{\it{o}}^*) \rm{d}\it A^* = \rm 0. \label{eqn:equilibrium}
\end{equation}
Substituting the second-order Taylor series expansion into equation \eqref{eqn:equilibrium} yields the following expression for the ice floe rotation rate normalized by the averaged ocean vorticity:
\begin{equation}
    \frac{\Omega_f}{\overline{\omega}_o/2} = \frac{\displaystyle \left[1- \frac{\pi^2}{8} \left(\frac{\gamma^2}{3}\right) + \frac{\pi^4}{64} \left( \frac{\gamma^4}{48} \right) \right]} {\displaystyle \left[1- \frac{\pi^2}{8} \left(\frac{\gamma^2}{2}\right) + \frac{\pi^4}{64} \left( \frac{\gamma^4}{24}\right) \right]}, \label{eq:TG_rotavg_approx}
\end{equation}
where $\gamma = R_f/R_e$.
Given that the ocean vorticity at the center-of-mass of the centered floe in the TG vortex is $\omega_{o,C} = 2A\pi^2/L_{TG}^2$, the ice floe rotation rate normalized by the ocean vorticity at the floe center-of-mass can be expressed as:
\begin{equation}
    \frac{\Omega_f}{\omega_{o,C}/2} = \left[1- \frac{\pi^2}{8} \left(\frac{\gamma^2}{3}\right) + \frac{\pi^4}{64} \left( \frac{\gamma^4}{48}\right) \right]. \label{eq:TG_rotCOM_approx}
\end{equation}
Equation \eqref{eq:TG_rotavg_approx} is in good agreement with the simulation results for the rotation rates normalized by the averaged ocean vorticity provided $R_f/R_e \leq 0.7$ (figure \ref{Fig:TG_cen_rot}$a$). 
In this range, the Taylor series approximation effectively restores the original functions.
However, we observed noticeable deviations in the normalized rotation rate when the ice floe size exceeded the eddy size due to limitations of the Taylor series expansion near the edge of the vortex cell. Equation \eqref{eq:TG_rotCOM_approx} aligns well with simulation results for the rotation rate normalized by the ocean vorticity at the floe center-of-mass for different floe-eddy size ratios because the ocean vorticity remains constant (figure \ref{Fig:TG_cen_rot}$b$).

We derived another analytical relation for the normalized rotation rate of the centered ice floe by approximating the floe shape as a square with the same characteristic length scale. 
With the equilibrium condition for ice-ocean stress torques, the analytical relation can be expressed as:
\begin{equation}
    \frac{\Omega_f}{\overline{\omega}_o/2} = \frac{12}{\pi^2\gamma^2} \left[1 - \frac{\pi \gamma}{2} \rm{cot} \it \left(\frac{\pi \gamma}{\rm{2}} \right)\right]. \label{eq:TG_rotavg_square}
\end{equation}
We positioned square-shaped floes centered with respect to the origin. This simplification allows for the direct use of trigonometric functions. As a result, the ice floe rotation rate normalized by the ocean vorticity at the floe center of mass can be written as:
\begin{equation}
    \frac{\Omega_f}{\omega_{o,C}/2} = \frac{\displaystyle 12}{\displaystyle \pi^2\gamma^2} \left[\frac{\displaystyle \rm{sin}\left(\it{\frac{\pi\gamma}{2}}\right)}{\displaystyle \frac{\pi\gamma}{2}} \right]^2  \left[1 - \frac{\pi\gamma}{2} \rm{cot} \it \left(\frac{\pi\gamma}{\rm{2}}\right)\right]. \label{eq:TG_rotCOM_square}
\end{equation}
Equation \eqref{eq:TG_rotavg_square} closely matches the simulation results for the rotation rates normalized by the averaged ocean vorticity because it incorporates trigonometric functions fully in the solution (figure \ref{Fig:TG_cen_rot}$a$).
However, equation \eqref{eq:TG_rotCOM_square} deviates from the simulation results for the rotation rates normalized by the center-of-mass ocean vorticity due to the square-shape approximation (figure \ref{Fig:TG_cen_rot}$b$). 
It is worth noting that, for the centered ice floes, the derived analytical relations \eqref{eq:TG_rotavg_approx}--\eqref{eq:TG_rotCOM_square} are functions solely of the floe-eddy size ratios. 

We also examined a more realistic scenario of an ice floe being positioned off-center with respect to the TG vortex cell. We conducted simulations with over 2,000 randomly distributed ice floes for each floe-eddy size ratio.
While centered ice floes remain at the vortex center, off-centered ice floes have the potential to drift away, depending on their radial positions. 
For the present analysis, only trapped ice floes were considered. 

\begin{figure}
  \centerline{\includegraphics[width=\textwidth]{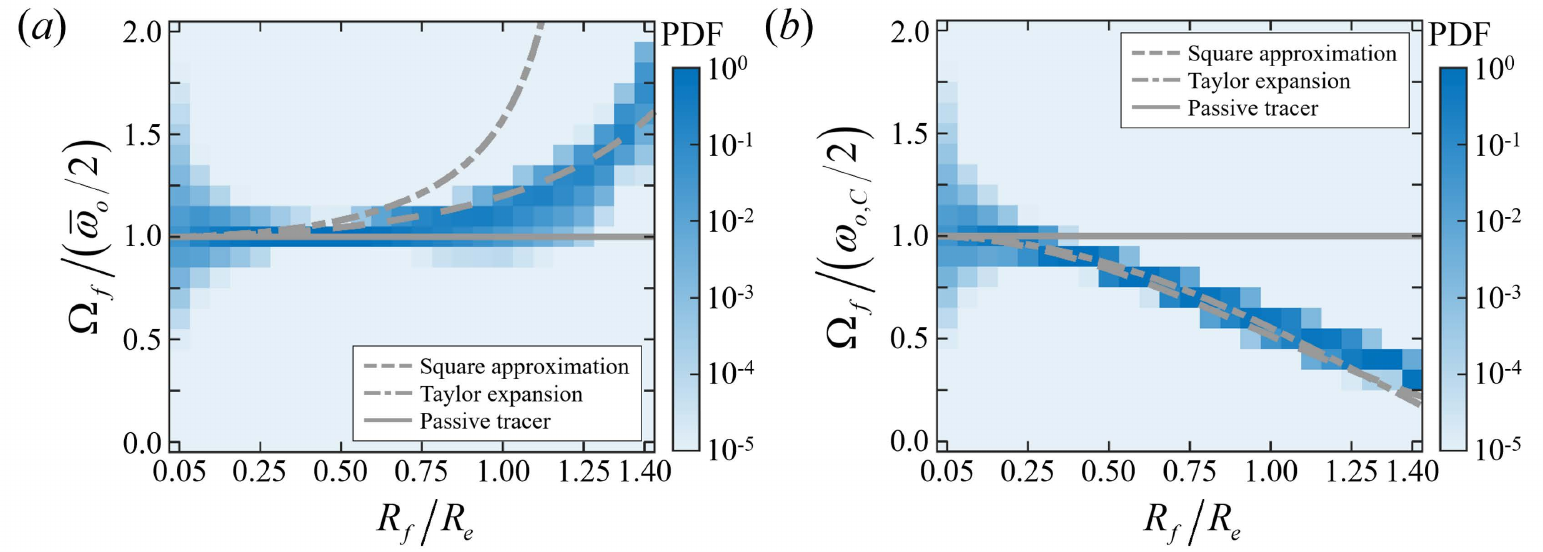}}
  \caption{Rotational motion of off-centered ice floes in a TG vortex field. PDF of ice floe rotation rates, $\Omega_f$, normalized by the ($a$) ocean vorticity averaged over the floe area, $\overline{\omega}_o$, and the ($b$) ocean vorticity at the floe center-of-mass, $\omega_{o,C}$, for different floe-eddy size ratios. The simulation results are compared to analytical relations using square-shape approximation (dashed lines) and Taylor series expansion (dashed-dotted lines), and the passive tracer case (solid lines).} \label{Fig:TG_rot_PDF}
\end{figure}

The normalized rotation rates of off-centered floes are dependent on the position of the floes relative to the vortex core. Therefore, to analyze this case, we computed probability density functions (PDF) over all floe positions for each bin of floe-eddy  size ratios (figure \ref{Fig:TG_rot_PDF}).
As discussed in the single floe analyses (\S\ref{subsec:results-TG}), the observed variability arises from floe inertia, causing a delay in ice floe response to changes in the underlying ocean eddy field.
This delay results in a discrepancy between the rotation of the ice floes and the underlying ocean eddies. 
Considering ice floe rotation rates normalized by the averaged ocean vorticity, ice floes behave as passive tracers provided $R_f/R_e \leq 0.7$ (figure \ref{Fig:TG_rot_PDF}$a$).
However, as the floe-eddy size ratio increases and approaches 1.4, the peaks in the normalized rotation rates shift to greater values, reaching approximately 1.8, similar to the case of centered ice floes (figure \ref{Fig:TG_cen_rot}$a$).
Around $R_f/R_e \approx 0.5$, the distribution narrows and exhibits higher peaks, suggesting that a solid body rotation approximation accurately captures floe motion within this range of floe-eddy size ratios.
This behavior can be attributed to the ice floe area filtering out ice floe-ocean stress, thereby reducing ice floe responsiveness to any variability within the vortex cell. 
As floe-eddy size ratios increase, excessive filtering of ocean information over the ice floe diminishes sensitivity to underlying ocean characteristics, resulting in a broadening of the PDF.
For ice floe rotation rates normalized by the ocean vorticity at the floe center-of-mass, ice floes behave as passive tracers when $R_f/R_e \leq 0.25$. 
The peaks of the PDF shift to smaller values as the size ratios increase, reaching approximately 0.25 at $R_f/R_e=1.4$ (figure \ref{Fig:TG_rot_PDF}$b$). 
Similar to the behavior observed for averaged ocean vorticity, the distributions are more dispersed for small floes, while clear peaks are evident for $R_f/R_e \geq 0.25$, as the center-of-mass ocean vorticity for larger floes tends to exceed the area-averaged vorticity.

Following the same approach as for centered ice floes, we derived analytical relations for the off-centered floes. 
By employing the Taylor series expansion, the analytical relations for the ice floe rotation rate normalized by the averaged ocean vorticity can be derived as:
\begin{equation}   
    \frac{\Omega_f}{\overline{\omega}_o/2} = \frac{\displaystyle \left\{1- \frac{\pi^2}{8} \left[\frac{\gamma^2}{3} + \left(\frac{r_f}{R_e}\right)^2 \right] + \frac{\pi^4}{64} \left[ \frac{\gamma^4}{48} + \frac{\gamma^2}{6}\left(\frac{r_f}{R_e}\right)^2 + \left(\frac{x_f}{R_e}\right)^2\left(\frac{y_f}{R_e}\right)^2\right] \right\}} {\displaystyle \left\{1- \frac{\pi^2}{8} \left[\frac{\gamma^2}{2} + \left(\frac{r_f}{R_e}\right)^2 \right] + \frac{\pi^4}{64} \left[ \frac{\gamma^4}{24} + \frac{\gamma^2}{4}\left(\frac{r_f}{R_e}\right)^2 + \left(\frac{x_f}{R_e}\right)^2\left(\frac{y_f}{R_e}\right)^2\right] \right\}},
    \label{eq:TG_offrotavg_approx}
\end{equation}
where $x_f$ and $y_f$ are the horizontal Cartesian coordinates of the floe center of mass, respectively, and $r_f = \sqrt{x_f^2 + y_f^2}$ is its radial position.
Similarly, the ice floe rotation rate normalized by the center-of-mass ocean vorticity can be then expressed as:
\begin{equation}
    \frac{\Omega_f}{\omega_{o,c}/2} = \frac{\displaystyle \left\{1- \frac{\pi^2}{8} \left[\frac{\gamma^2}{3} + \left(\frac{r_f}{R_e}\right)^2 \right] + \frac{\pi^4}{64} \left[ \frac{\gamma^4}{48} + \frac{\gamma^2}{6}\left(\frac{r_f}{R_e}\right)^2 + \left(\frac{x_f}{R_e}\right)^2\left(\frac{y_f}{R_e}\right)^2\right] \right\}} {\displaystyle \left[1- \frac{\pi^2}{8} \left(\frac{r_f}{R_e}\right)^2 + \frac{\pi^4}{64} \left(\frac{x_f}{R_e}\right)^2\left(\frac{y_f}{R_e}\right)^2 \right]}.
    \label{eq:TG_offrotCOM_approx}
\end{equation}
In contrast to the centered ice floe cases, relations \eqref{eq:TG_offrotavg_approx} and \eqref{eq:TG_offrotCOM_approx} not only depend on the floe-eddy size ratio, but also on the radial positions of the ice floes. Therefore, the normalized rotation rates were averaged over all $r_f$ values for each of the floe-eddy size ratio bins. As $r_f \rightarrow 0$, the analytical relations for the off-centered floes (equations \ref{eq:TG_offrotavg_approx} and \ref{eq:TG_offrotCOM_approx}) converge to the relations for the centered floes (equations \ref{eq:TG_rotavg_approx} and \ref{eq:TG_rotCOM_approx}). Equation \eqref{eq:TG_offrotavg_approx} is in good agreement with the PDF peaks for small floe-eddy size ratios (figure \ref{Fig:TG_rot_PDF}$a$). Their differences become evident around $R_f/R_e \approx 0.5$ and become more pronounced for larger size ratios due to the issues of the Taylor series expansion near the cell edges. Conversely, equation \eqref{eq:TG_offrotCOM_approx} closely aligns with the PDF peaks of ice floe rotation rates normalized by the center-of-mass vorticity for most floe-eddy size ratios (figure \ref{Fig:TG_rot_PDF}$b$). 

The square-shape approximation for off-centered floes yields the same analytical relations as for centered floes. Thus, equations \ref{eq:TG_rotavg_square} and \ref{eq:TG_rotCOM_square} are also used to describe the rotation rates of off-centered floes.
These equations match the PDF peaks over a wide range of floe-eddy size ratios and only show minor deviations at ratios around $R_f/R_e=1.4$ (figures \ref{Fig:TG_rot_PDF}$a$ and \ref{Fig:TG_rot_PDF}$b$). 
Overall, the square-shape approximation provides accurate estimates of the PDF peak compared to the Taylor series approximation.


\begin{figure}
  \centerline{\includegraphics[width=\textwidth]{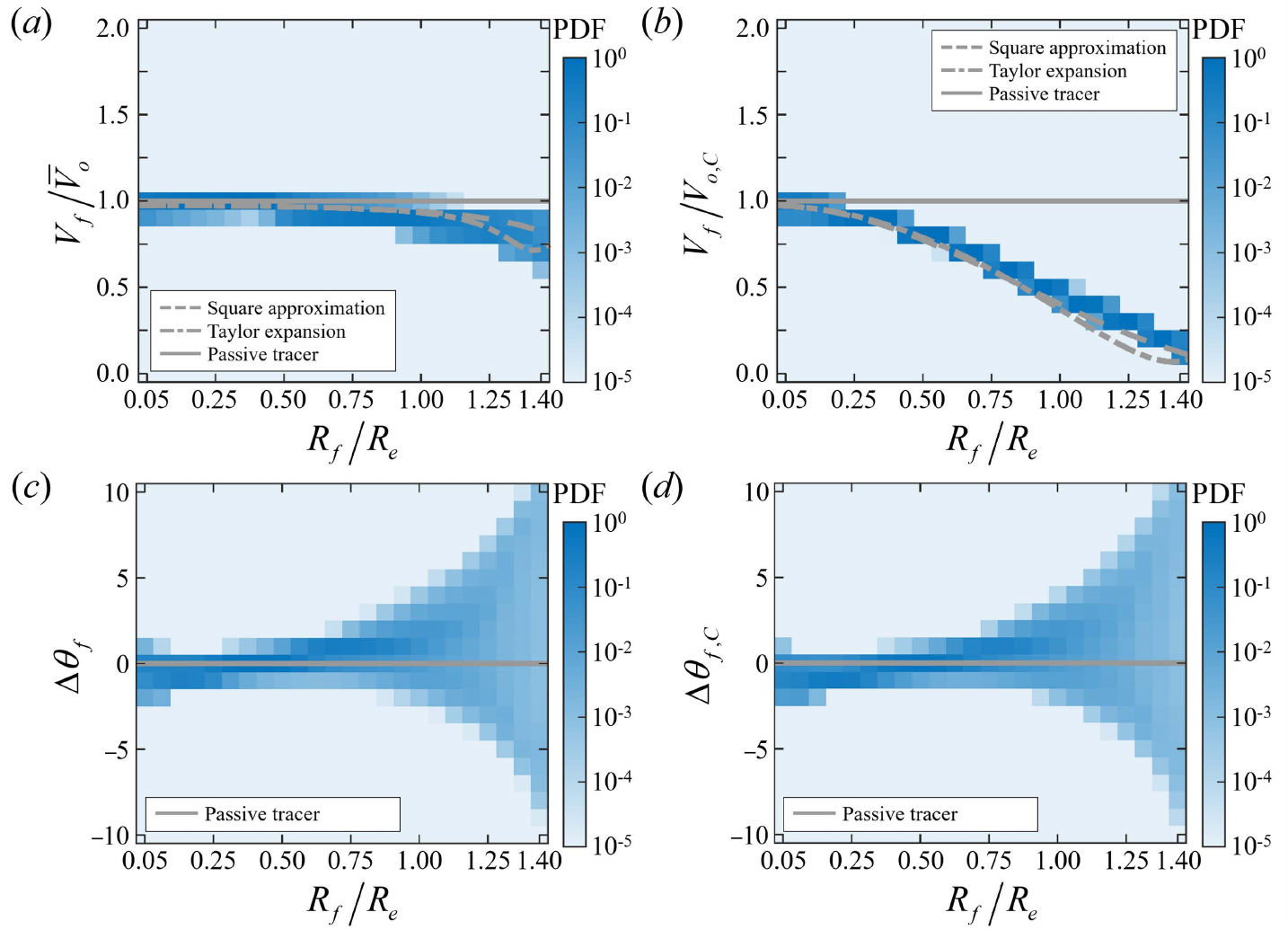}}
  \caption{Translational motion of off-centered ice floes in a TG vortex field. PDF of ice floe speed, $V_f$, normalized by the ($a$) ocean speed averaged over the floe area, $\overline{V}_o$, and the ($b$) ocean speed at the floe center-of-mass, $V_{o,C}$, and ($c$ and $d$) angle deviations. For these last quantities, ice and ocean velocity angles are calculated using the direction of each velocity vector with respect to the positive $x$-axis, measured in the counter-clockwise direction. Then, deviations are computed by subtracting the angle of ($c$) the averaged ocean velocity, $\Delta \theta_f$, and the ($d$) center-of-mass ocean velocity, $\Delta \theta_{f,C}$, from that of the ice floe velocity, for different floe-eddy size ratios. The simulation results are compared to analytical relations using square-shape approximation (dashed lines) and Taylor series expansion (dashed-dotted lines), and the passive tracer case (solid lines). } \label{Fig:TG_vel_ang_PDF}
\end{figure}


The PDFs of normalized ice floe speeds and floe-ocean angle deviations depend on the floe-eddy size ratios (figure \ref{Fig:TG_vel_ang_PDF}), corresponding to the results of single floe analyses depicted in figure \ref{Fig:T_F_time_evol}.
For the ice floe speed normalized by the ocean speed averaged over the floe area, the PDF peaks at unity ($V_f/\overline{V}_o = 1$), resembling the behavior of a passive tracer (figure \ref{Fig:TG_vel_ang_PDF}$a$).
As the floe-eddy size ratio increases and approaches 1.4, the normalized speed at the PDF peaks gradually decreases to $V_f/\overline{V}_o = 0.75$.
In contrast to the ice floe rotation rate results, the PDFs of normalized floe speeds for small-sized floes exhibit narrow distributions.
This trend can be attributed to the fact that the Coriolis force and the pressure gradient force due to the sea surface tilt depend on the velocity difference, $\Delta \bf{u}_{\it{f}} = \overline{\bf{u}}_{\it{o}}-\bf{u}_{\it{f}}$, while the torque generated by these forces depends solely on ocean velocity, $\overline{\bf{u}}_{\it{o}}$, with no contribution from the Coriolis force. As a result, floe speed is less responsive to changes in ocean speed. 
This is reflected in the third term of equation \eqref{eq:sea-ice-eqns_force_non}, which is derived using the geostrophic balance between the pressure gradient force due to sea surface tilt and the Coriolis force.

Overall, the ocean speed at the floe center-of-mass is greater than the ocean speed averaged over the floe area. The peaks of the PDF for the floe speed normalized by the center-of-mass ocean speed decrease to 0.25 as the floe-eddy size ratio increases to 1.4 (figure \ref{Fig:TG_vel_ang_PDF}$b$).
It is useful to note that the velocity difference between the ice floe and the underlying ocean remains relatively constant across different floe-eddy size ratios, such that $\lvert \Delta \bf{u}_{\it{f}} \rvert \it \approx C_{\rm{0}}$, where $C_0$ denotes a constant.
This value can be derived by equating the magnitude of the ice floe-ocean stress (the first term on the right-hand side of equation \ref{eq:sea-ice-eqns_force_non}) to a combination of the Coriolis force on the ice floe and the pressure gradient force due to the sea surface tilt (the third term  on the right-hand side of equation \ref{eq:sea-ice-eqns_force_non}) under torque equilibrium, as follows: $\lvert \Delta \bf{u}_{\it{f}}^{\it{*}} \it \rvert^2 \it /H_{f,o}^* \sim \lvert \rm{\Delta} \bf{u}_{\it{f}}^{\it{*}} \it \rvert /Ro$. 
Then, the floe speed normalized by the averaged ocean velocity can be expressed as:
\begin{equation} \label{eq:TG_offvelavg}
    \frac{V_f}{\overline{V}_{o}} \approx 1 - \frac{C_0}{\overline{V}_{o}},
\end{equation}
implying that for lower averaged ocean speeds as in the case of larger floes, there is a reduction in normalized floe speeds.


Similar to equations \eqref{eq:TG_offrotavg_approx}--\eqref{eq:TG_offrotCOM_approx}, the analytical relation for ice floe speed normalized by the ocean speed averaged over the floe area and by the ocean speed at the floe center-of-mass can be derived using the Taylor series expansion and the square-shape approximation. 
In relation \eqref{eq:TG_offvelavg}, the averaged ocean speed can be approximated using a Taylor series expansion as follows:
\begin{equation}
    \overline{V}_o = \left(\frac{\pi}{2}\right)^2 \left(\frac{A_{TG}}{R_e}\right) \left(\frac{r_f}{R_e}\right) M_o, \label{eq:TG_offocnvelavg_Taylor}
\end{equation}
where $M_o=[(x_f/r_f)^2M_{o,1}^2 + (y_f/r_f)^2M_{o,2}^2]^{1/2}$. $M_{o,1}$ and $M_{o,2}$ are given by
\begin{subequations}
\begin{align}
\begin{split}
    M_{o,1} & = 1 - \frac{1}{12}\left(\frac{\pi}{2}\right)^2 \left\{3\gamma^2 + 2\left[\left(\frac{x_f}{R_e}\right)^2 + 3\left(\frac{y_f}{R_e}\right)^2\right]\right\} \\ 
    & + \frac{1}{96} \left(\frac{\pi}{2}\right)^4 \left\{\gamma^4 + 2\gamma^2 \left[\left(\frac{x_f}{R_e}\right)^2 + 3\left(\frac{y_f}{R_e}\right)^2\right] + 8\left(\frac{x_f}{R_e}\right)^2\left(\frac{y_f}{R_e}\right)^2 \right\},\label{eq:M_o1}
\end{split}
\end{align}
\begin{align}
\begin{split}
    M_{o,2} & = 1 - \frac{1}{12}\left(\frac{\pi}{2}\right)^2 \left\{3\gamma^2 + 2\left[3\left(\frac{x_f}{R_e}\right)^2 + \left(\frac{y_f}{R_e}\right)^2\right]\right\} \\ 
    & + \frac{1}{96} \left(\frac{\pi}{2}\right)^4 \left\{\gamma^4 + 2\gamma^2 \left[3\left(\frac{x_f}{R_e}\right)^2 + \left(\frac{y_f}{R_e}\right)^2\right] + 8\left(\frac{x_f}{R_e}\right)^2\left(\frac{y_f}{R_e}\right)^2 \right\}.\label{eq:M_o2}
\end{split}
\end{align}
\end{subequations}
Similarly, the ice floe speed normalized by the center-of-mass ocean speed can be expressed using the averaged ocean speed as follows:
\begin{equation} \label{eq:TG_offvelCOM}
    \frac{V_f}{V_{o,C}} = \left(1 - \frac{C_0}{\overline{V}_{o}}\right)\left( \frac{\overline{V}_{o}}{V_{o,C}} \right).
\end{equation}
The center-of-mass ocean speed can also 
be approximated using the Taylor series expansion:
\begin{equation} 
    V_{o,C} = \left(\frac{\pi}{2}\right)^2 \left(\frac{A_{TG}}{R_e}\right) \left(\frac{r_f}{R_e}\right) M_{o,C}, \label{eq:TG_offocnvelCOM_Taylor}
\end{equation}
where $M_{o,C}$ is given by
\begin{align}
\begin{split}
    M_{o,C}^2 & = 1 - \frac{1}{3}\left(\frac{\pi}{2}\right)^2\left(\frac{r_f}{R_e}\right)^2 \left[1 + 4\left(\frac{x_f}{r_f}\right)^2\left(\frac{y_f}{r_f}\right)^2 \right] \left[1 + \frac{1}{12}\left(\frac{\pi}{2}\right)^4\left(\frac{x_f}{R_e}\right)^2\left(\frac{y_f}{R_e}\right)^2\right]  \\ 
    & + \frac{1}{36}\left(\frac{\pi}{2}\right)^4\left(\frac{r_f}{R_e}\right)^4 \left[1 + 18\left(\frac{x_f}{r_f}\right)^2\left(\frac{y_f}{r_f}\right)^2 + \frac{1}{4}\left(\frac{\pi}{2}\right)^4\left(\frac{r_f}{R_e}\right)^4\left(\frac{x_f}{r_f}\right)^4\left(\frac{y_f}{r_f}\right)^4 \right]. \label{eq:M_oC}
\end{split}
\end{align}
Note that for comparison with the simulation results, the normalized floe speeds in the analytical relations were averaged over all $r_f$ values for each bin of floe-eddy size ratios. 
Similarly, using the square-shape approximation, the averaged ocean speeds can be approximated as the following:
\begin{align}
\begin{split}
    \overline{V}_o = \left(\frac{\pi}{2}\right) & \left(\frac{A_{TG}}{R_e}\right) \left[\mathrm{sin}\left(\frac{\pi\gamma}{2}\right) \bigg/ \left(\frac{\pi\gamma}{2}\right) \right]^2 \\
    & \left[ \mathrm{cos}^2 \left(\frac{\pi}{2} \frac{x_f}{R_e}\right) \mathrm{sin}^2\left(\frac{\pi}{2} \frac{y_f}{R_e}\right) + \mathrm{sin}^2 \left(\frac{\pi}{2} \frac{x_f}{R_e}\right) \mathrm{cos}^2 \left(\frac{\pi}{2} \frac{y_f}{R_e}\right) \right]^{1/2}. \label{eq:TG_offocnvelavg_sq}
\end{split}
\end{align}
The center-of-mass ocean speed can also be obtained as:
\begin{equation}
    V_{o,C} = \left(\frac{\pi}{2}\right) \left(\frac{A_{TG}}{R_e}\right) \left[\mathrm{cos}^2 \left(\frac{\pi}{2} \frac{x_f}{R_e}\right) \mathrm{sin}^2 \left(\frac{\pi}{2} \frac{y_f}{R_e}\right) + \mathrm{sin}^2 \left(\frac{\pi}{2} \frac{x_f}{R_e}\right) \mathrm{cos}^2 \left(\frac{\pi}{2} \frac{y_f}{R_e}\right) \right]^{1/2}. \label{eq:TG_offocnvelCOM_sq}
\end{equation}

The analytical relation using the square-shape approximation (equations \ref{eq:TG_offocnvelavg_sq} and \ref{eq:TG_offocnvelCOM_sq}) shows good agreement with the peaks of the normalized speeds, while the analytical relation using the Taylor series expansion (equations \ref{eq:TG_offocnvelavg_Taylor} and \ref{eq:TG_offocnvelCOM_Taylor}) exhibits deviations for $R_f/R_e \geq 1$ (figures \ref{Fig:TG_vel_ang_PDF}$a$ and \ref{Fig:TG_vel_ang_PDF}$b$). 
The constant $C_0=0.15$ was chosen, producing the best-fit to the PDF peaks.
Note that as $r_f \rightarrow 0$, the analytical relations for the normalized floe speed (equations \ref{eq:TG_offvelavg} and \ref{eq:TG_offvelCOM}) converge to the centered floe case as expected.

The PDFs of the angle deviation between ice floe velocity and ocean velocity are also contingent on the floe-eddy size ratio (figures \ref{Fig:TG_vel_ang_PDF}$c$ and \ref{Fig:TG_vel_ang_PDF}$d$).
As discussed in the single floe analysis (figure \ref{Fig:T_F_time_evol}), the velocity of larger floes exhibits greater deviations from ocean velocity, including the averaged ocean velocity and the center-of-mass velocity, thus resulting in more dispersed distributions for larger floe-eddy size ratios.
While the vectors of the averaged ocean velocity and the center-of-mass ocean velocity differ slightly in magnitude, their orientations are the same. 
This observation can be readily confirmed using the square-shape approximation, as follows:
\begin{equation}
    \mathrm{tan}\overline{\theta}_o = - \mathrm{tan}\left(\frac{\pi}{2}\frac{x_f}{R_e}\right) \bigg/ \mathrm{tan}\left(\frac{\pi}{2}\frac{y_f}{R_e}\right) = \mathrm{tan}\theta_{o,C}, \label{eq:ocn_angle}
\end{equation}
indicating that the angle of the averaged ocean velocity is equal to the angle of the center-of-mass ocean velocity.

In addition to calculating ocean velocity and vorticity beneath single ice floes, we can also compute local ocean vorticity from Green's theorem by using ice floe trajectory information. Passive tracers trapped in an eddy exhibit closed-loop trajectories. In this case, we can estimate the averaged ocean vorticity of the region enclosed by the trajectory as $\tilde{\omega}_o = \int \mathbf{u}_{tr} \cdot \mathrm{d}\mathbf{S}_{tr}$, where $\mathbf{u}_{tr}$ is the tracer velocity, $\mathbf{S}_{tr}$ is the tracer trajectory, and $\sim$ denotes the averaged quantity over the closed region. 
Here, since $\mathbf{u}_{tr} = \mathbf{u}_{o}$, the true ocean vorticity, $\tilde{\omega}_o$, can be obtained. Similarly, trapped ice floes have been observed to form nearly closed-loop patterns, with the endpoints slightly offset from the starting points. However, as previously discussed, the velocity of ice floes corresponds to the filtered ocean velocity over the floe area but is not exactly the same. As a result, any averaged ocean vorticity estimate within a trajectory-enclosed region, $\tilde{\omega}_f$, does not match the true averaged ocean vorticity of the same region, $\tilde{\omega}_f \neq \tilde{\omega}_o$. Nonetheless, their ratios can be approximated using the ice floe and underlying ocean velocities as:
\begin{equation}
    \frac{\tilde{\omega}_f}{\tilde{\omega}_o} = \frac{\oint \mathbf{u}_{f} \cdot \mathrm{d}\mathbf{S}_{f}} {\oint \mathbf{u}_{o,C} \cdot \mathrm{d}\mathbf{S}_{f}} \approx \frac{V_f}{V_{o,C}}, \label{eq:TG_offregionvor}
\end{equation}
where $\mathbf{u}_{o,C}$ is the ocean velocity at the floe center of mass and $\mathbf{S}_f$ is the ice floe trajectory. 

Floe trajectories are not perfectly closed, with just a short segment between the start and endpoints. Hence, the line integration in equation \ref{eq:TG_offregionvor} is conducted from the first to the last points. The true averaged ocean vorticity can be calculated by integrating the center-of-mass ocean velocity along the floe trajectory. The ratio of the ocean vorticity estimate based on ice floe velocity to the true ocean vorticity, $\tilde{\omega}_f / \tilde{\omega}_o$, can be approximated by the normalized ice floe speed, $V_f/V_{o,C}$ (equation \ref{eq:TG_offregionvor}).

\begin{figure}
  \centerline{\includegraphics[width=0.5\textwidth]{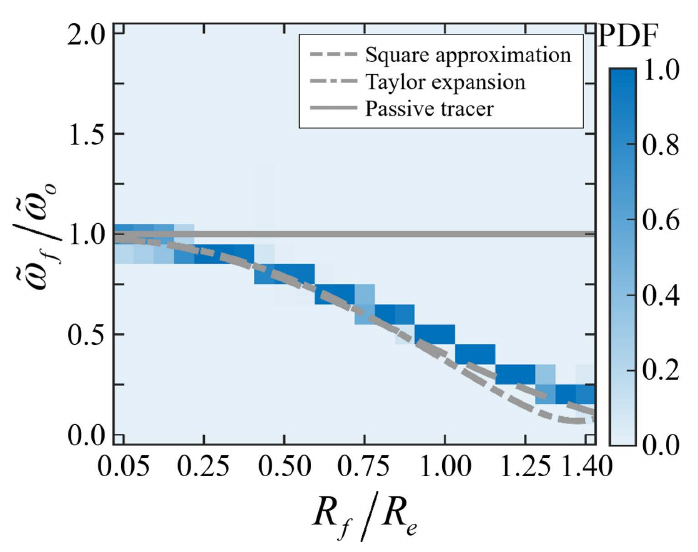}}
  \caption{Averaged ocean vorticity estimates from ice floe trajectories in a TG vortex field. PDF of ratios between the averaged ocean vorticity of the region enclosed by ice floe trajectories, $\tilde{\omega}_f$, and the true averaged ocean vorticity of the region, $\tilde{\omega}_o$, for different floe-eddy size ratios. The simulation results are compared to analytical relations using the square-shape approximation (dashed line) and Taylor series expansion (dashed-dotted line), and the passive tracer case (solid line).} \label{Fig:TG_cir_PDF}
\end{figure}

The PDFs of the ratio between the ocean vorticity estimate and the true ocean vorticity depend on floe-eddy size ratios (figure \ref{Fig:TG_cir_PDF}).
The floe speed normalized by the center-of-mass ocean speed significantly decreases for larger floe-eddy size ratios (figure \ref{Fig:TG_vel_ang_PDF}$b$), resulting in smaller vorticity ratios.
Furthermore, the peaks of the PDF for most floe-eddy size ratios align with the analytical relation for normalized floe speed using the square-shape approximation (equations \ref{eq:TG_offvelCOM}, \ref{eq:TG_offocnvelavg_sq}, and \ref{eq:TG_offocnvelCOM_sq}). 
The relation using the Taylor series expansion (equations \ref{eq:TG_offvelCOM}--\ref{eq:M_oC}) is in good agreement with the PDF peaks for $R_f/R_e \leq 0.75$, showing a discrepancy for larger ratios due to the coverage limit of the Taylor series expansion near the edge of the vortex cell.

\subsection{Statistics of ice floe kinematics in a QG flow field}  \label{subsec:results-QG}

\begin{figure}
  \centerline{\includegraphics[width=\textwidth]{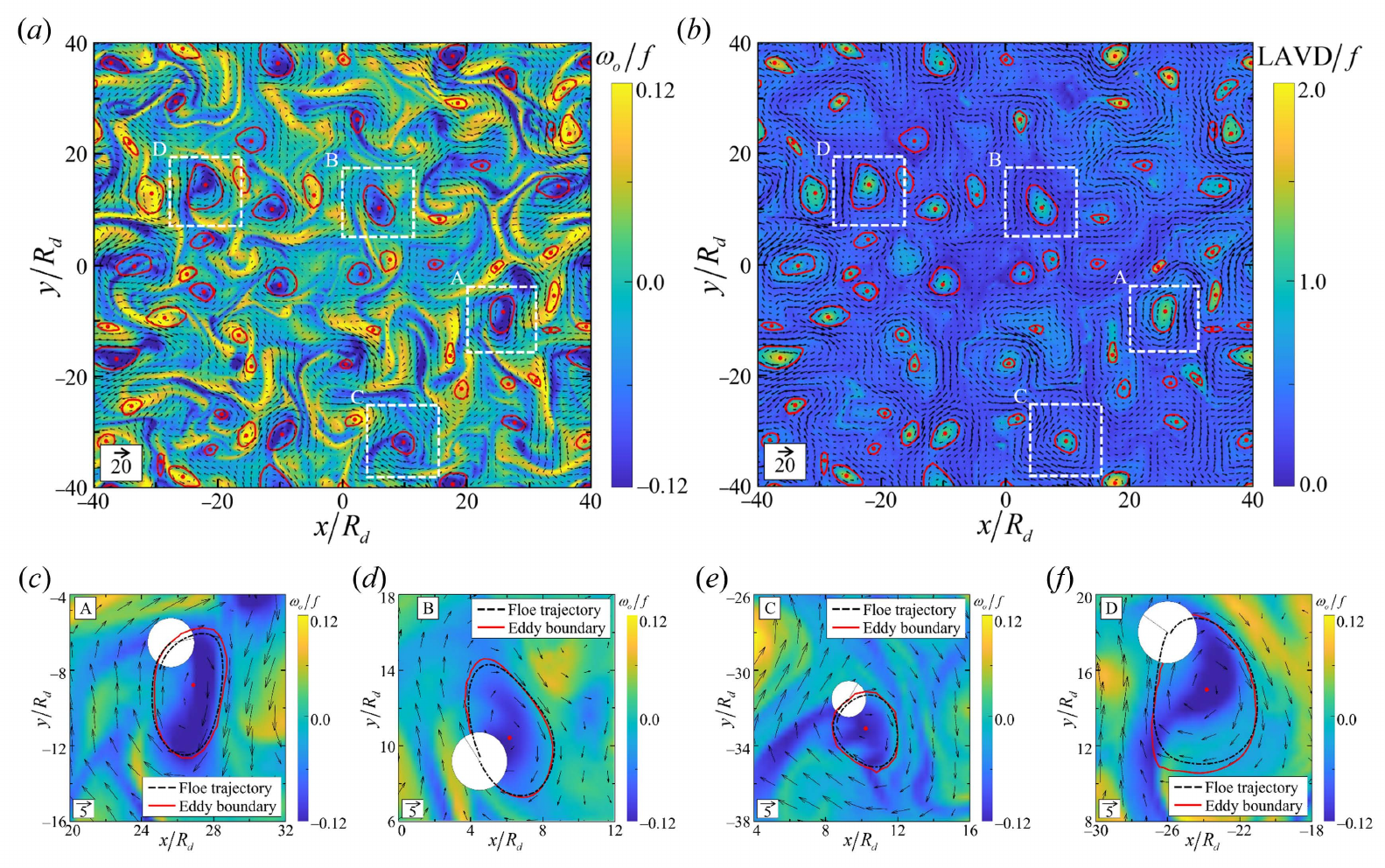}}
  \caption{Motion of ice floes in a QG flow field. Boundaries (red lines) and centers (red dots) of the detected eddies are shown in the field of ($a$) ocean vorticity and ($b$) LAVD normalized by the Coriolis parameter $f$. ($c$--$f$) Representative cases of trajectories (black dashed lines) of trapped ice floes within local eddies marked as $A$ to $D$ with white dashed boxes in the QG and LAVD flow fields, respectively. The color and arrows in the figure correspond to the magnitude of the normalized vorticity and LAVD and the magnitude and direction of the velocity at a given location, respectively. The floes (white circle) are positioned at the endpoint of their trajectories.} \label{Fig:QG_eddy_traj}
\end{figure}

We performed simulations of ice floes in a QG flow field to apply the derived analytical relations to a more realistic ocean eddy field (figure \ref{Fig:QG_eddy_traj}).
Passive tracers were released into the flow field, and ocean vorticity was interpolated to the Lagrangian tracer positions.
Then, a LAVD-based eddy detection method \citep{Haller2016} was used to identify the boundaries of eddies by searching for the outermost closed contour of the LAVD, which indicates local rotation relative to mean rotation. The LAVD was computed by averaging the vorticity deviation along the Lagrangian tracer trajectory as follows:
\begin{equation}
    \mathrm{LAVD} \left(x_0,y_0\right) = \omega_o' \left[X\left(x_0,y_0\right),Y\left(x_0,y_0\right)\right],
    \label{eq:LAVD-form}
\end{equation}
where $(X,Y)$ are the coordinates of tracers with an initial position of $(x_0,y_0)$, and $\omega_o'$ is the vorticity deviation from the spatial average over the whole domain. 
Given that the ocean field is time-independent in the present study, the time component is not considered in equation \eqref{eq:LAVD-form}. A total of 70 eddies were identified in the QG flow field (figure \ref{Fig:QG_eddy_traj}$a$);
their morphologies become more apparent in the LAVD field (figure \ref{Fig:QG_eddy_traj}$b$). Note that the LAVD has been widely used to capture vorticity-dominated structures in atmospheric and oceanic flows and has also been used as a baseline for evaluating trajectory-based diagnostics to quantify the kinematics of an underlying fluid \citep{Aksamit2024}. By leveraging the rotational relationship between floes and underlying eddies, the LAVD has the potential to be applied to the trajectories and rotation rates of ice floes for eddy detection purposes.

We selected four varying-size representative eddy cases to examine the kinematic link between drifting ice floes and the underlying flow field (figures \ref{Fig:QG_eddy_traj}$c$--$f$). To this end, we randomly released over 2,000 ice floes near each eddy, identifying floes trapped within vortex cores (Appendix \ref{sec:app_eddy_identification}). Ice floes released near the eddy boundary exhibited nearly closed-loop trajectories resembling the shapes of the eddies (figures \ref{Fig:QG_eddy_traj}$c$--$f$). Hence, the trajectories of individual trapped ice floes provide a direct estimate of local eddy length scales, whereby the size of the largest enclosed region by ice floe trajectories, $R_{traj}=\sqrt{A_{traj}/\pi}$, was chosen as a trajectory-derived length scale for each local eddy. 
Here, $A_{traj}$ is the area of the largest enclosed region by ice floe trajectories. This length scale closely matches the eddy size in the TG vortex.

In QG eddies, ice floes simultaneously covering both the inside and outside regions of an eddy often become entrapped by the eddy, resulting in minor discrepancies between eddy size and trajectory-derived length scales.
These differences arise due to variations in vorticity and velocity distributions among different QG eddies, leading to diverse trajectory shapes and trajectory-derived length scales. 
In the selected QG eddies, $R_f/R_{traj}$ range from 0.05 to 1.25. Beyond these ranges, ice floes generally remain translating near the eddies or drift away from them over the simulation period.

\begin{figure}
  \centerline{\includegraphics[width=\textwidth]{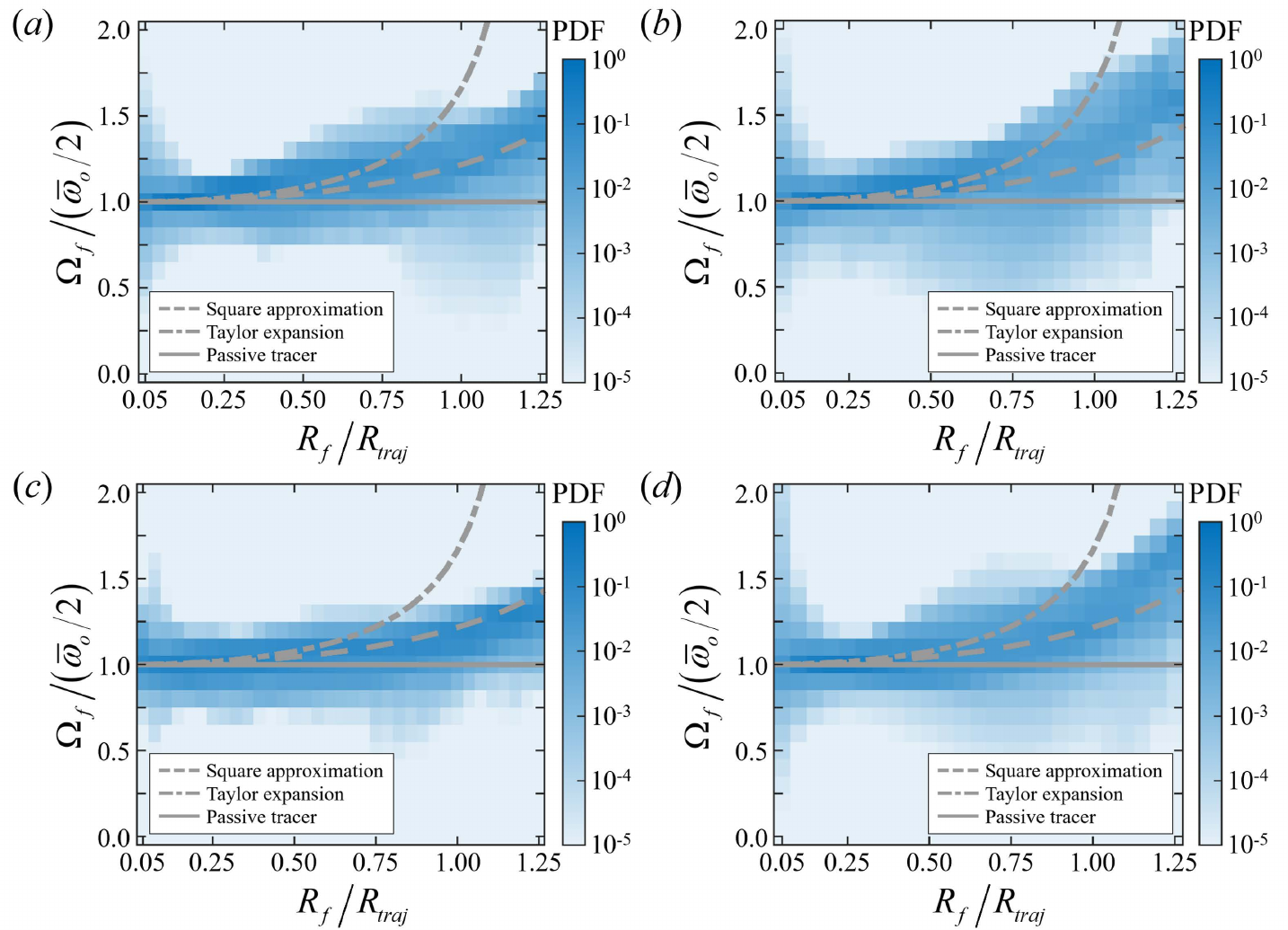}}
  \caption{Rotational motion of trapped ice floes in local QG eddies. ($a$--$d$) PDF of ice floe rotation rates, $\Omega_f$, normalized by the averaged ocean vorticity over the floe area, $\overline{\omega}_o$, for different ratios between floe size and trajectory-derived length scale, $R_f/R_{traj}$, in local eddies marked as $A$ to $D$ with white dashed boxes in a QG flow field, respectively. The simulation results are compared to analytical solutions using square-shape approximation (dashed line), Taylor series expansion (dashed-dotted line), and the passive tracer case (solid line).} \label{Fig:QG_rotavg}
\end{figure}

We investigated the rotational relationship between trapped ice floes and the underlying local eddies for the four eddies marked in figure \ref{Fig:QG_eddy_traj}. 
Despite variations in eddy size and shape, the normalized rotation rates of ice floes showed similar qualitative and quantitative trends to those in the idealized vortex cases (figure \ref{Fig:QG_rotavg}). Notably, the PDFs of ice floe rotation rates, normalized by the ocean vorticity averaged over the floe area, peak at unity when $R_f/R_{traj} \leq 0.5$ for all four cases (figure \ref{Fig:QG_rotavg}). Note that the distributions of the normalized floe rotation rates in the QG eddy cases are more dispersed than in the TG vortex cases. We attribute this discrepancy to the highly nonlinear, deformed eddies dominant in the QG flow field. The PDF peaks become more pronounced in the QG cases as the floe size approaches half of the trajectory-derived length scale ($R_f/R_{traj} = 0.5$). The peaks shift toward 1.5 as $R_f/R_{traj}$ hits 1.25. Analytical relations derived for the TG vortex effectively characterize this behavior. The relation from the Taylor series expansion (equation \ref{eq:TG_offrotavg_approx}) aligns with PDF peaks for smaller $R_f/R_{traj}$, while the relation from the square-shape approximation (equation \ref{eq:TG_rotavg_square}) agrees well with the peaks across most size ratios.

\begin{figure}
  \centerline{\includegraphics[width=\textwidth]{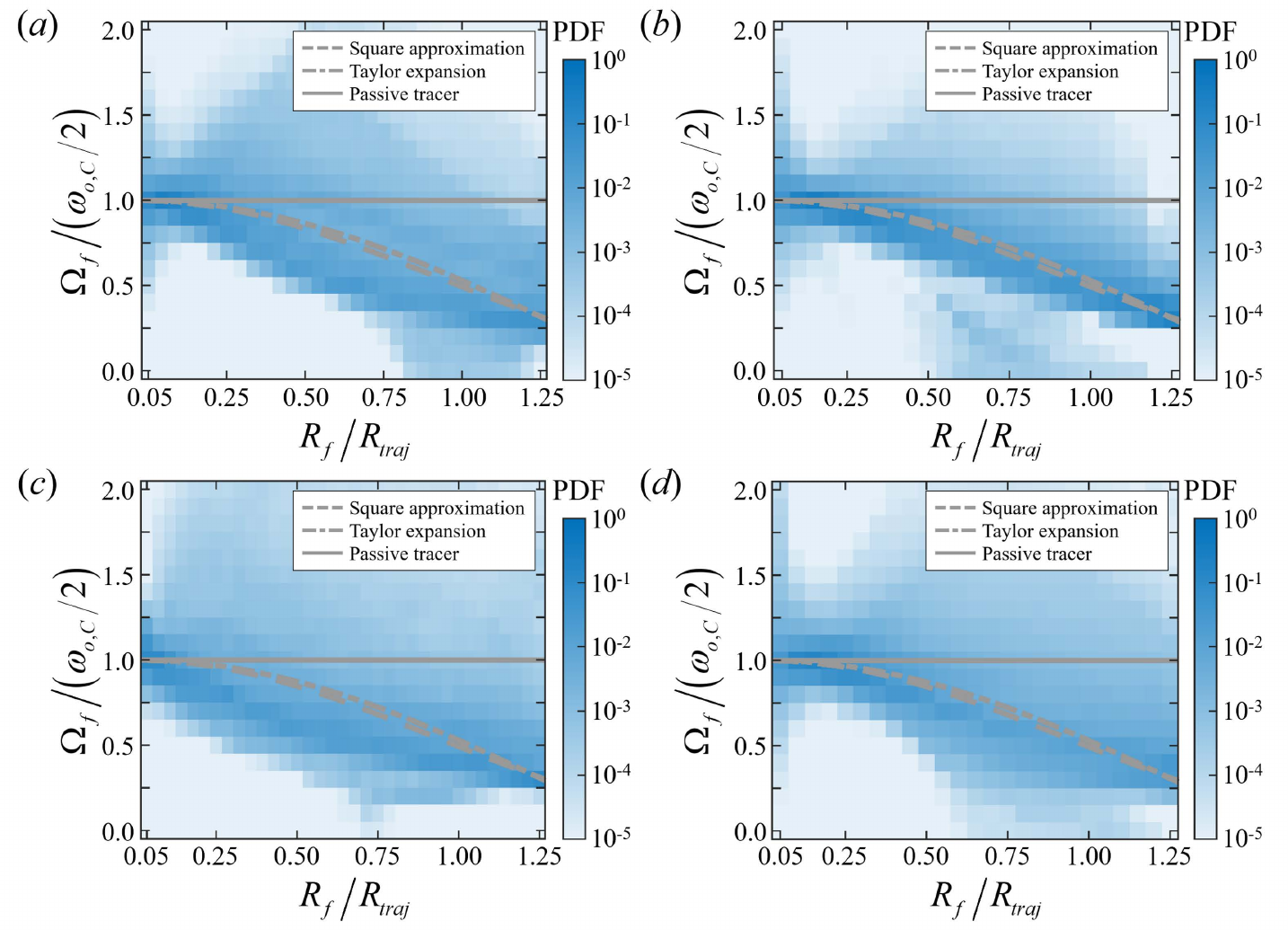}}
  \caption{Rotational motion of trapped ice floes in local QG eddies. ($a$--$d$) PDF of ice floe rotation rates, $\Omega_f$, normalized by the ocean vorticity at the floe center of mass, $\omega_{o,C}$, for different ratios between floe size and trajectory-derived length scale in local eddies marked as $A$ to $D$ with white dashed boxes in a QG flow field, respectively. The simulation results are compared to analytical solutions using square-shape approximation (dashed line), Taylor series expansion (dashed-dotted line), and the passive tracer case (solid line).} \label{Fig:QG_rotCOM}
\end{figure}

The PDFs of ice floe rotation rates normalized by the ocean vorticity at the floe center of mass exhibit similar qualitative behaviors compared to the idealized vortex case (figure \ref{Fig:QG_rotCOM}).
As $R_f/R_{traj}$ increases to 1.25, the PDF peaks decrease to 0.25, consistent with the TG vortex cases (figure \ref{Fig:TG_rot_PDF}$b$).
However, analytical relations \eqref{eq:TG_rotCOM_square} and \eqref{eq:TG_offrotCOM_approx}, derived for the TG vortex, exhibit discrepancies with the PDF peaks due to differences in the eddy structures of the flow fields.
Vorticities near the center of QG eddies exhibit smaller spatial gradients compared to the TG vortex, leading to smaller normalized rotation rates in the QG eddy cases.
In addition, the center-of-mass ocean vorticity changes abruptly across regions, contrasting with the averaged ocean vorticity, such that the rotation rates normalized by the center-of-mass ocean vorticity show more discrepancies with the TG vortex cases and analytical relations derived for them.

\begin{figure}
  \centerline{\includegraphics[width=\textwidth]{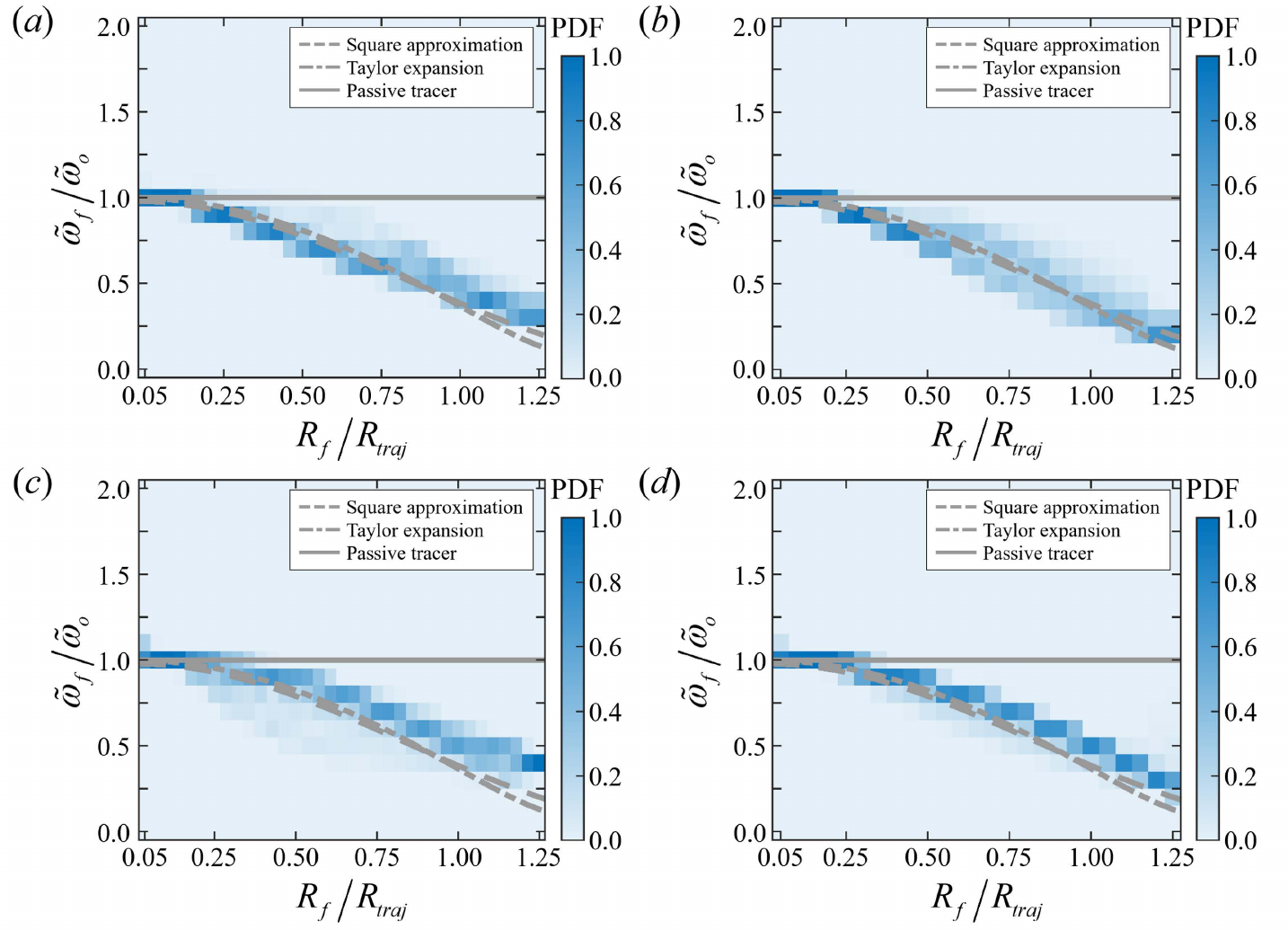}}
  \caption{Averaged ocean vorticity estimate using ice floe trajectories in local QG eddies. ($a$--$d$) PDF of ratios between averaged ocean vorticity of the floe trajectory-enclosed region, $\tilde{\omega}_f$, normalized by true averaged ocean vorticity of the same region, $\tilde{\omega}_o$, for different ratios between floe size and trajectory-derived length scale in local eddies marked as $A$ to $D$ with white dashed boxes in a QG flow field, respectively. The simulation results are compared to analytical solutions using square-shape approximation (dashed line), Taylor series expansion (dashed-dotted line), and the passive tracer case (solid line).} \label{Fig:QG_cir}
\end{figure}

The ocean vorticity estimate of the region enclosed by ice floe trajectories in QG eddies exhibits similar trends to the estimate in the TG vortex (figure \ref{Fig:QG_cir}). 
The PDFs of the ratio between the ocean vorticity estimate and true ocean vorticity reach their peaks at unity.
However, these peaks decrease to 0.25 for larger $R_f/R_{traj}$, reaching up to 1.25. 
This trend is likely due to a greater discrepancy between the averaged ice floe velocity and the ocean velocity at the floe center of mass in QG eddies.
Overall, analytical relation \eqref{eq:TG_offregionvor} captures the qualitative behavior of the PDF peaks, albeit with minor quantitative discrepancies observed for certain $R_f/R_{traj}$. 

The observed ice floe motions and their analytical relations to the ocean, derived from the idealized eddies, offer insights into the rotational relationship between ice floes and local eddies in the QG flow field.
These findings demonstrate the potential applicability of these derived relations to analyze ocean eddies and estimate their vorticity using ice floes. 
The complexity in the QG case stems from how each eddy exhibits a unique structure, velocity, and vorticity distribution, as illustrated in figure \ref{Fig:QG_eddy_traj}. These fundamental differences between the TG vortex and QG eddies limit the direct application of these relationships to QG fields. Nonetheless, further investigation, including analyzing the effects of deformed eddy configurations and different vortex types, holds promise for developing a comprehensive framework to characterize the ocean eddy field from ice floe satellite remote sensing observations.

\section{Discussion and further analyses} \label{sec:discussion}

Following the analysis with idealized free-drifting ice floes, we investigated additional relevant factors influencing the link between ice floe rotation and the vorticity of underlying ocean eddies. Specifically, we discuss the effects of ice floe thickness, atmospheric winds, and floe-floe collisions, corresponding to specific sea ice concentrations, on the rotational relationship between ice floes and the TG vortex. 

\subsection{Effects of ice floe thickness on ice floe kinematics}  \label{subsec:diss-thickness}

\begin{figure}
  \centerline{\includegraphics[width=\textwidth]{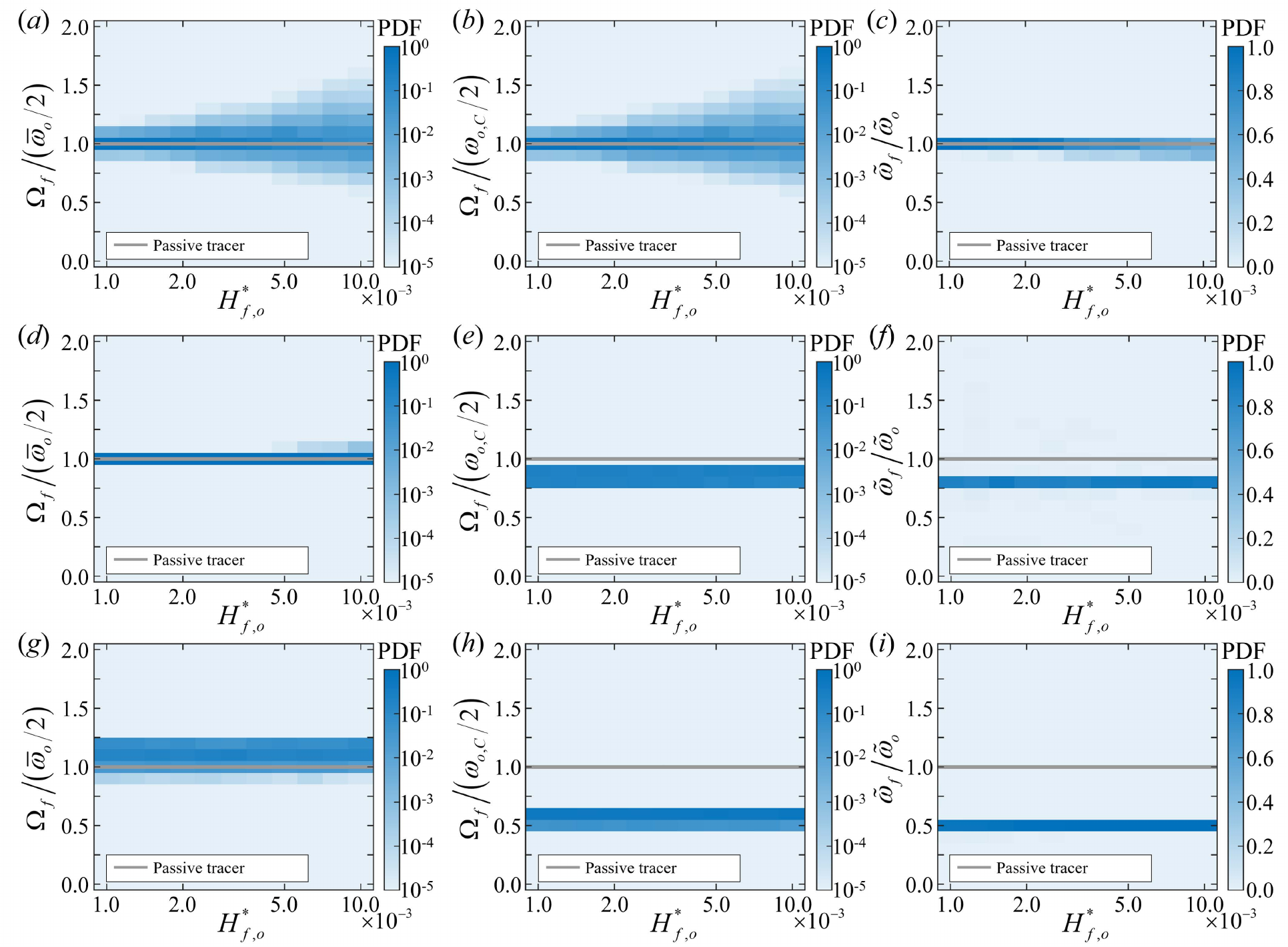}}
  \caption{Motion of trapped ice floes with different thicknesses in a TG vortex. PDF of ice floe rotation rate, $\Omega_f$, normalized by the ($a,d,g$) averaged ocean vorticity over the floe area, $\overline{\omega}_o$, and the ($b,e,h$) ocean vorticity at the floe center-of-mass, $\omega_{o,C}$. and of ($g,h,i$) ratio between estimated ocean vorticity averaged over the trajectory-enclosed region, $\tilde{\omega}_f$, and true ocean vorticity of the same region, $\tilde{\omega}_o$, for different floe-eddy size ratios, $R_f/R_e = $ ($a,b,c$) 0.1, ($d,e,f$) 0.5. and ($g,h,i$) 1.0.} \label{Fig:TG_thick_rot}
\end{figure}

The motion of ice floes is affected by their inertia, which is closely linked to their thickness. Thicker ice floes inherently possess greater inertia than thinner floes, leading to increased discrepancies between ice floe motions and the underlying ocean kinematics. Observations of ice floe thickness in the BG MIZ are limited \citep{Haas2009}, but estimates typically fall within the order of $O(0.1)$ m during the spring-to-summer season \citep{Manucharyan2022b}, especially for young and first-year ice, as estimated based on upward-looking sonars \citep{Krishfield2014}. To assess the influence of thickness on our results, we performed additional simulations considering floes with ice floe thicknesses ranging from 0.1 m to 1 m, corresponding to $H_{f,o}^* = 1.0 \times 10^{-3}$--$1.0 \times 10^{-2}$, for three different floe-eddy size ratios. We examined the PDFs of normalized ice floe rotation rates and the trajectory-derived ocean vorticity estimates. 

The sensitivity of the rotational relationship to ice floe thickness depends on the floe-eddy size ratio, with smaller ice floes being the most responsive to changes in ice floe thickness (figure \ref{Fig:TG_thick_rot}). 
For $R_f/R_e=0.1$, the PDF of the normalized rotation rates consistently peaks at unity (figures \ref{Fig:TG_thick_rot}$a$ and \ref{Fig:TG_thick_rot}$b$). 
However, as ice floes become thicker, their inertia increases, leading to greater deviations in their rotation from the ocean rotation. 
As a result, the peaks gradually diminish, and the distributions become more dispersed.
In addition, thicker ice floes tend to follow a more pronounced spiral trajectory, resulting in skewed distributions toward larger normalized rotation rates.
Similarly, the PDFs of the ratio between averaged ocean vorticity estimates of the trajectory-enclosed region and true averaged ocean vorticity peak at unity (figure \ref{Fig:TG_thick_rot}$c$).
However, for thicker ice floes, the peaks gradually decrease, and the distributions become more dispersed and skewed toward lower ratios.

As the floe-eddy size ratio increases, the effect of ice floe thickness on the rotational relationship becomes relatively minor (figures \ref{Fig:TG_thick_rot}$d-i$).
Specifically, at $R_f/R_e=0.5$, thicker ice floes exhibit more dispersed distributions for the rotation rate normalized by the averaged ocean vorticity, skewed toward larger values (figure \ref{Fig:TG_thick_rot}$d$).
In contrast, these floes show negligible changes in the rotation rate normalized by the center-of-mass vorticity and in the averaged  ocean vorticity estimates normalized by the true ocean vorticity (figures \ref{Fig:TG_thick_rot}$e$ and \ref{Fig:TG_thick_rot}$f$).
At $R_f/R_e=1.0$, the PDFs become more dispersed compared to floes with $R_f/R_e=0.5$ (figures \ref{Fig:TG_thick_rot}$g-i$), as discussed in \S\ref{sec:results}. 
However, thicker ice floes exhibit nearly identical distributions to thinner ice floes, indicating negligible effects of ice floe thickness.
Overall, thicker ice floes tend to follow more pronounced spiral trajectories due to their increased inertia.
Nevertheless, these larger floes are less sensitive to changes in ocean information over the floe area, as this information is filtered out, resulting in minor changes in the rotational relationship between ice floes and the underlying ocean.

\subsection{Effects of atmospheric winds on ice floe kinematics}  \label{subsec:diss-wind}

While our analysis primarily focused on sea ice-ocean interactions, surface wind drag also influences ice floe motions in practical scenarios. 
Strong winds predominantly exert force on ice floes, potentially weakening the kinematic relationship to ocean vorticity and increasing uncertainty when inferring this information.
Here, we investigate the effect of atmospheric winds on the rotational relationship between ice floes and the underlying ocean eddies by incorporating the surface wind drag term in equations \eqref{eq:sea-ice-eqns_force_non} and \eqref{eq:sea-ice-eqns_torque_non} into our analysis.

In general, atmospheric winds have larger length scales compared to upper-ocean eddies, potentially causing ice floes to trace straight trajectories rather than curved ones \citep{Lopez2021,Manucharyan2022b}.
Thus, we considered unidirectional atmospheric winds with consistent positive speeds across the entire domain in both the zonal and meridional directions. These winds do not directly affect the rotation of ice floes under homogeneous conditions (i.e., uniform surface roughness and thickness within floes). However, unidirectional winds influence the translational motion of the ice floes, leading to increased discrepancies between their rotations and the rotations of the underlying ocean.

We investigated the motion of ice floes under weak and strong wind conditions. In low winds, ice floes tend to remain within the vortex cell. In these cases, surface wind stress to surface ocean stress ratios, $\tau_{a,ref}/\tau_{o,ref}$, range from 0 to 0.1, and wind velocities have magnitudes from 0 to 2.5 m/s. Here, the stress ratio is calculated relative to zero ice floe speed. As wind speeds increase, the influence of wind forcing on ice floe motion becomes more significant, leading to ice floes escaping the vortex cell and a notable decrease in the number of trapped ice floes, especially the large floes. In high winds, stress ratios range from 0.2 to 2.0, and wind speeds vary from 2.5 to 10.6 m/s, corresponding to observed wind speeds in the MIZ \citep{Kozlov2022}.

\begin{figure}
  \centerline{\includegraphics[width=\textwidth]{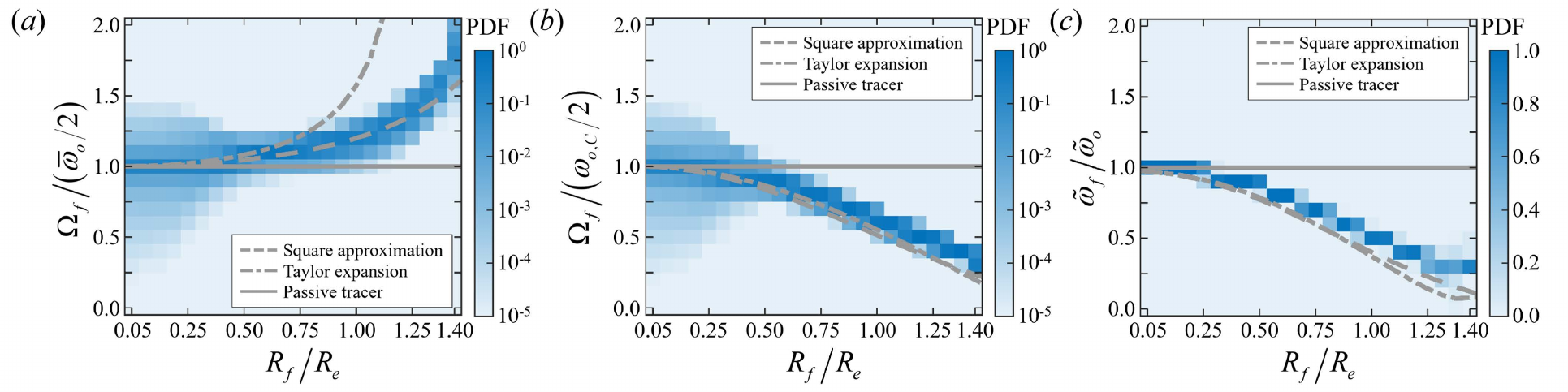}}
  \caption{Motion of trapped ice floes in a TG vortex with low winds. PDF of ice floe rotation rate, $\Omega_f$, normalized by the ($a$) averaged ocean vorticity over the floe area, $\overline{\omega}_o$, and the ($b$) ocean vorticity at the floe center of mass, $\omega_{o,C}$, and ($c$) of the ratio between averaged ocean vorticity estimate of the trajectory-enclosed region, $\tilde{\omega}_f$, and true averaged ocean vorticity of the same region, $\tilde{\omega}_o$, for different floe-eddy size ratios. The wind stress to ocean stress ratio is $\tau_a/\tau_o = 0.10$. The simulation results are compared to analytical solutions using square-shape approximation (dashed line) and Taylor series expansion (dashed-dotted line), and the passive tracer case (solid line).} \label{Fig:TG_wind_Re}
\end{figure}

In low wind scenarios ($\tau_{a,ref}/\tau_{o,ref}=0.1$), the PDFs of the normalized rotation rates and the ratios between averaged ocean vorticity estimates and true ocean vorticities exhibit similar peaks and distributions (figure \ref{Fig:TG_wind_Re}) compared to those under zero wind conditions (figures \ref{Fig:TG_rot_PDF} and \ref{Fig:TG_cir_PDF}).
For normalized rotation rates, distributions become more dispersed for most floe-eddy size ratios, with peaks slightly shifting toward higher values due to  wind forcing deforming ice floe trajectories toward cell boundaries.
In addition, the analytical relations \eqref{eq:TG_rotavg_square}--\eqref{eq:TG_offrotCOM_approx} and \eqref{eq:TG_offregionvor} align well with PDF peaks, demonstrating the potential applicability of these derived relations under low wind conditions.

\begin{figure}
  \centerline{\includegraphics[width=\textwidth]{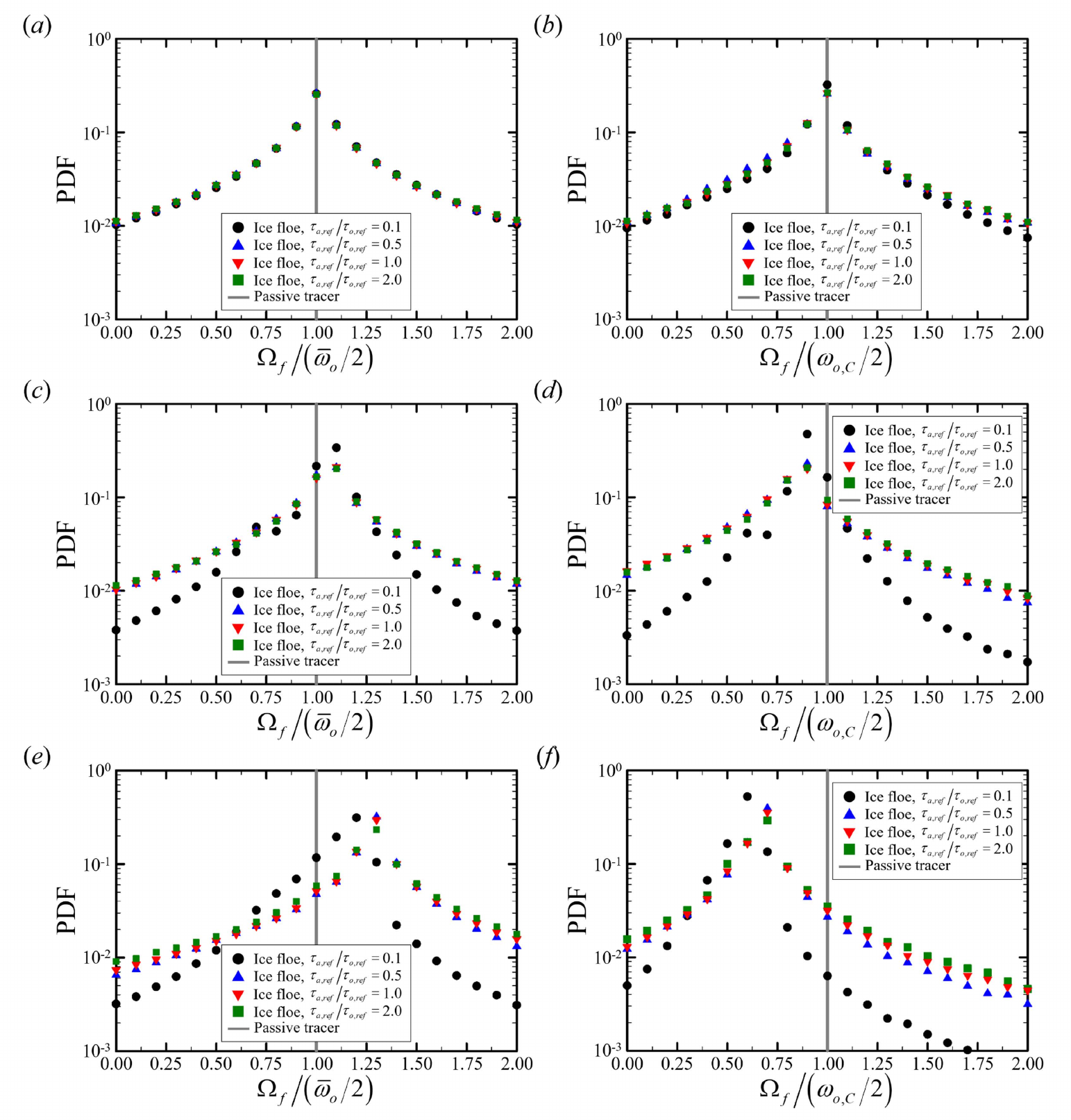}}
  \caption{Motion of trapped ice floes in a TG vortex with high winds. PDF of ice floe rotation rate, $\Omega_f$, normalized by the ($a,c,e$) averaged ocean vorticity over the floe area, $\overline{\omega}_o$, and the ($b,d,f$) ocean vorticity at the floe center of mass, $\omega_{o,C}$, for different wind stress to ocean stress ratios, $\tau_{a,ref}/\tau_{o,ref} = $ $0.1$ (black circle), $0.5$ (blue up-pointing triangle), $1.0$ (red down-pointing triangle), $2.0$ (green square). The floe-eddy size ratios are $R_f/R_e$ = ($a,b$) 0.1, ($c,d$) 0.5, ($e,f$) 1.0. The simulation results are compared to the passive tracer case (solid line).} \label{Fig:TG_wind_strong}
\end{figure}

In high winds, the effects of surface wind drag on the rotational relationship vary depending on the floe-eddy size ratios (figure \ref{Fig:TG_wind_strong}). Note that all ice floes are considered in the analyses. For $R_f/R_e=0.1$, the effects of wind forcing on the rotation rate normalized by the averaged ocean vorticity are negligible (figure \ref{Fig:TG_wind_strong}$a$). However, wind forcing marginally decreases the peaks of the PDFs and results in slightly more dispersed distributions (figure \ref{Fig:TG_wind_strong}$b$). As the floe-eddy size ratio increases, the effects of wind forcing become more pronounced, resulting in skewed distributions. For $R_f/R_e=0.5$ and $1.0$, the PDF peaks of rotation rates normalized by the averaged ocean vorticity shift toward greater values (figures \ref{Fig:TG_wind_strong}$c$ and \ref{Fig:TG_wind_strong}$e$), whereas the peaks of rotation rates normalized by the center-of-mass ocean vorticity shift toward lower values (figures \ref{Fig:TG_wind_strong}$d$ and \ref{Fig:TG_wind_strong}$f$).
The peak values in both cases decrease, and the distributions become more dispersed as the stress ratio increases from 0.1 to 0.5. Beyond this range, the PDFs begin to converge, showing minor differences in the distribution. The peaks slightly decrease, and the distributions become more dispersed at larger stress ratios.

Overall, the wind breaks off the influence of ocean eddies on ice floe rotation, with the effect being most pronounced under strong wind conditions. In such cases, ice floes tend to move around and along the boundaries of the vortex cell, where ocean and wind stresses are comparable, resulting in skewed distributions of normalized rotation rates. At the same stress ratios, larger floes show significant changes in the peaks and distributions of their rotation rates due to their greater coverage of the flow field. 


\subsection{Effects of floe-floe collisions on ice floe kinematics} \label{subsec:diss-collision}

Ice floes with no collisions are typically found in regions with low sea ice concentrations.
In contrast, in areas of higher sea ice concentrations, floe-floe collisions become more common due to the densely packed distribution of ice floes \citep{Lepparanta2011} affecting floe rotations \citep{Brenner2023}. 
These collisions exert contact forces that weaken the connection between ice floes and the underlying ocean, introducing noise into the estimation of ocean information through ice floe motions. 
In this section, we explored the effects of floe-floe collisions on the rotational relationship between ice floes and the underlying ocean to assess the feasibility of using ice floes with collisions for inferring ocean kinematics.

Simulations were conducted with 2,000 randomly released ice floes, and rotation rate measurements began once the overlapping areas between ice floes were reduced to less than 10$\%$ of their total area.
Sea ice concentration was calculated as the total ice floe area divided by the size of the smallest rectangular domain confining all ice floes over the simulation time.
Since the domain size is determined by the instantaneous positions of the ice floes, sea ice concentration can vary slightly over time. Thus, the time-averaged sea ice concentration was used to represent each simulation.
In equations \eqref{eq:sea-ice-eqns_force} and \eqref{eq:sea-ice-eqns_torque}, the contact forces, $\mathbf{F}_{j,k}$, and the corresponding torques, $\mathbf{r}_{j,k}' \times \mathbf{F}_{j,k}$, were incorporated into the simulations. The simulation parameters are consistent with those described for ice floes in the BG MIZ, as detailed in \cite{Manucharyan2022b} and summarized in table \ref{tab:sim_param}.
For the ice floe setup, the total of 2,000 ice floes was divided into smaller subsets to ensure that the number of collisions remained within specified ranges of interest while maintaining the desired sea ice concentration.
Similar to the strong wind cases (\S\ref{subsec:diss-wind}), the simulated ice floes exhibited significant deviations from the closed-loop patterns observed in the idealized cases due to collisions and thus all results were indiscriminately considered.

\begin{figure}
  \centerline{\includegraphics[width=\textwidth]{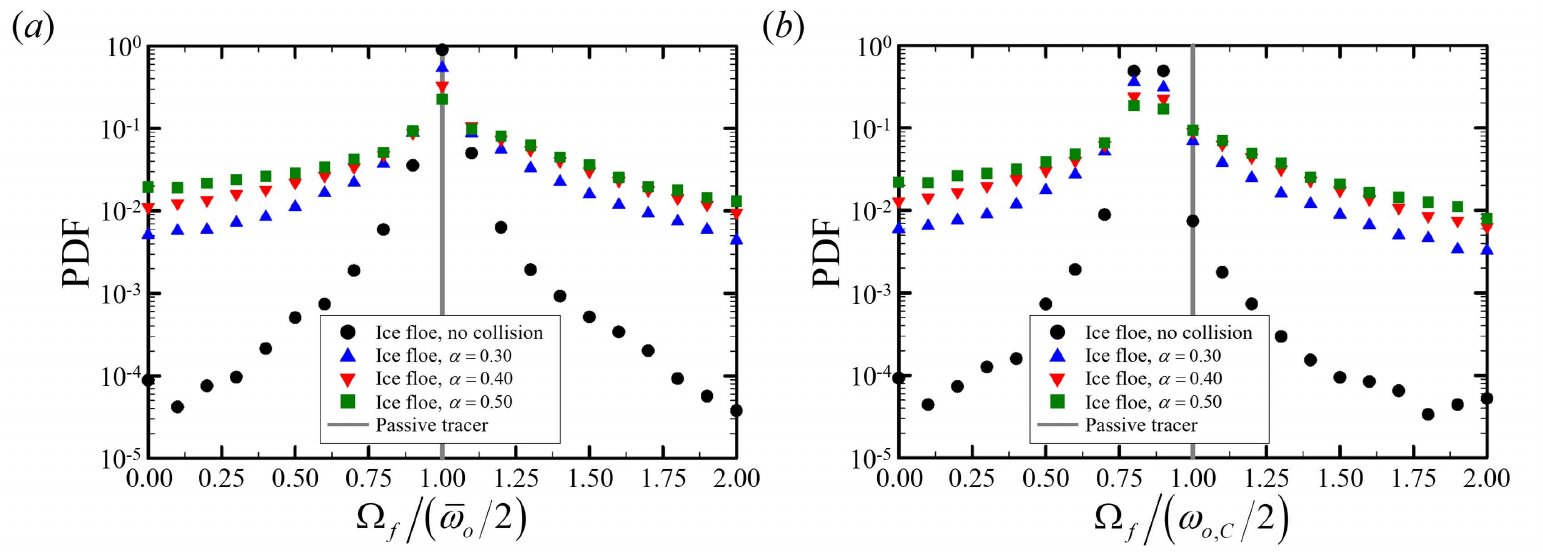}}
  \caption{Motion of ice floes with collisions in a TG vortex. PDF of ice floe rotation rate, $\Omega_f$, normalized by the ($a$) averaged ocean vorticity over the floe area, $\overline{\omega}_o$, and the ($b$) ocean vorticity at the floe center of mass, $\omega_{o,C}$, for different sea ice concentrations, $\alpha =$ 0.30 (blue up-pointing triangle), 0.40 (red down-pointing triangle), and 0.50 (green square). The floe-eddy size ratio is $R_f/R_e = 0.5$. The results are compared to the freely drifting floes (black circles) and the passive tracer case (solid lines).} \label{Fig:TG_col_alpha}
\end{figure}

We examined the PDFs of normalized ice floe rotation rates for different sea ice concentrations, $\alpha$, and compared them to floes with no collisions (figure \ref{Fig:TG_col_alpha}).
As a representative case, we selected floes with $R_f/R_e = 0.5$, showing a clear peak in the PDFs.
Sea ice concentrations ranging from $\alpha=0.3$ to $0.5$ were considered, reflecting typical moderate values in the MIZ, where concentrations range from $\alpha=0.15$ to $0.80$.
Overall, the PDF peaks occur at the same rotation rates for different concentrations, but higher concentrations lead to more dispersed distributions.
For rotation rates normalized by the averaged ocean vorticity, the PDF peak decreases from 0.89 (no collision case) to 0.22 ($\alpha=0.5$) (figure \ref{Fig:TG_col_alpha}$a$).
Similarly, for rotation rates normalized by the center-of-mass vorticity, the PDF peak decreases from 0.49 (no collision case) to 0.19 ($\alpha=0.5$) (figure \ref{Fig:TG_col_alpha}$b$).
These changes are caused by increasing collision counts at higher concentrations, which reduce the influence of ocean vorticity on ice floe rotation. The average collision count increases from zero (no-collision case) to 160 ($\alpha=0.5$).
Despite these changes, the distributions retain their shapes across different concentrations, as collisions show no directional bias in altering floe rotation.

\begin{figure}
  \centerline{\includegraphics[width=\textwidth]{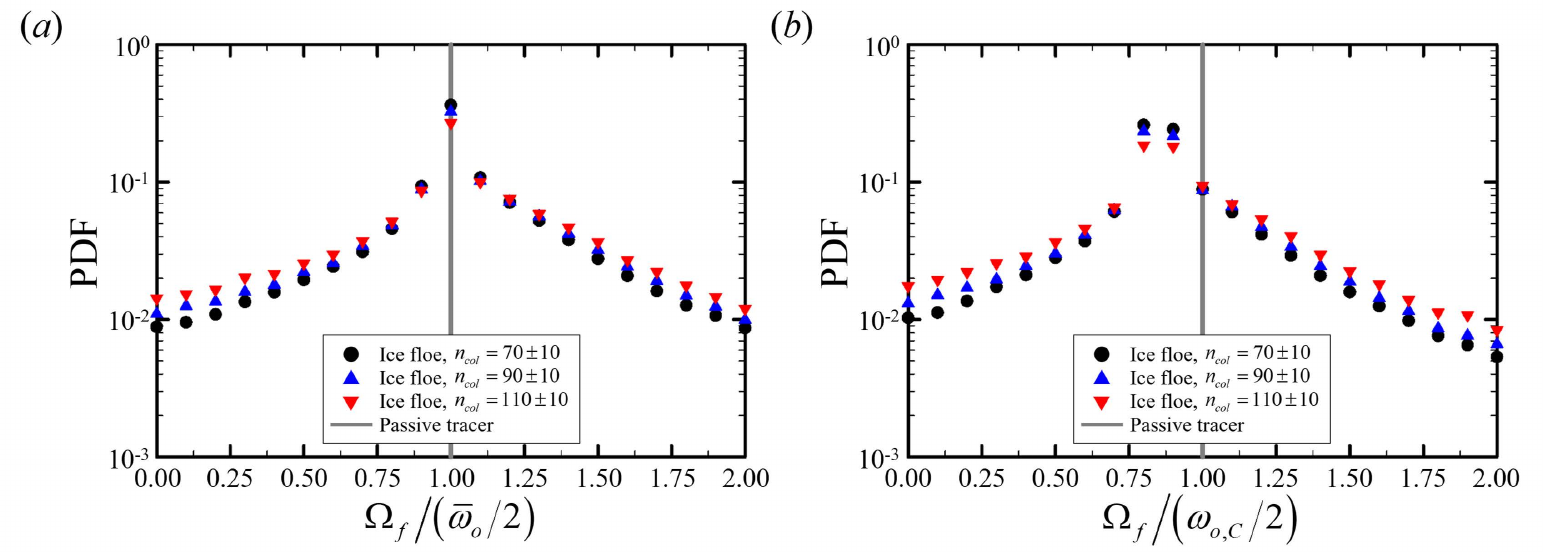}}
  \caption{Motion of ice floes with collisions in a TG vortex. PDF of ice floe rotation rate, $\Omega_f$, normalized by the ($a$) averaged ocean vorticity over the floe area, $\overline{\omega}_o$, and the ($b$) ocean vorticity at the floe center of mass, $\omega_{o,C}$ for different collision count ranges, $n_{col} =$ 70 $\pm$ 10 (black circle), 90 $\pm$ 10 (blue up-pointing triangle), and 110 $\pm$ 10 (red down-pointing triangle). The floe-eddy size ratio is $R_f/R_e = 0.5$. The results are compared to the passive tracer case (solid lines). } \label{Fig:TG_col_number}
\end{figure}

While sea ice concentration represents the overall fraction of ice floes in the domain, collision count reflects the extent of contact forces and corresponding torques affecting ice floe motions.
We compared the PDFs of normalized rotation rates for cases with different collision counts, $n_{col} = 70 \pm 10, 90 \pm 10, 110 \pm 10 $, at $R_f/R_e=0.5$ to assess the influence of floe-floe collisions (figure \ref{Fig:TG_col_number}). 
Within each simulation run, the collision counts vary over time and only the time ranges corresponding to the selected collision counts were considered for the analysis.
The PDFs consistently peak at the same rotation rates for different collision counts but become increasingly dispersed as collision counts rise.
This behavior aligns with the trends observed for different sea ice concentrations, where higher concentrations generally coincide with more frequent collisions, assuming constant floe sizes.

\begin{figure}
  \centerline{\includegraphics[width=\textwidth]{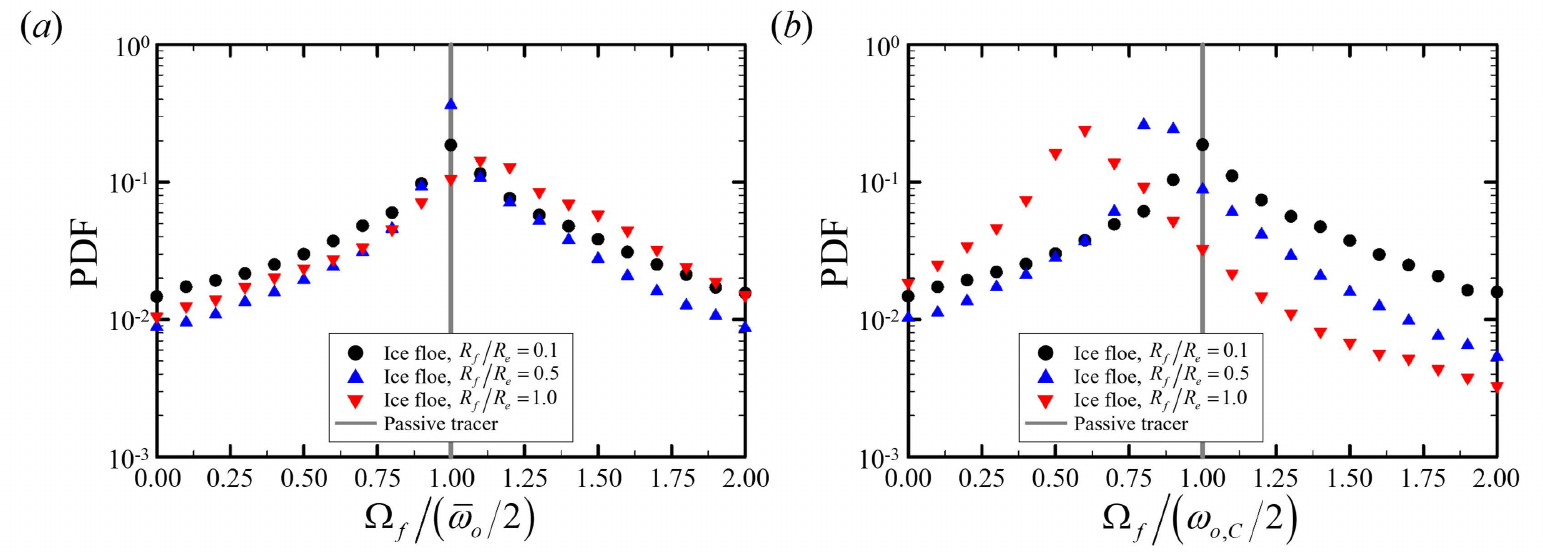}}
  \caption{Motion of ice floes with collisions in a TG vortex. PDF of ice floe rotation rate, $\Omega_f$, normalized by the ($a$) averaged ocean vorticity over the floe area, $\overline{\omega}_o$, and the ($b$) ocean vorticity at the floe center-of-mass, $\omega_{o,C}$ for different floe-eddy size ratios, $R_f/R_e = $ 0.1 (black circle), 0.5 (blue up-pointing triangle), and 1.0 (red down-pointing triangle). The collision count range is $n_{col} = 70 \pm 10$. The results are compared to the passive tracer case (solid lines).} \label{Fig:TG_col_size}
\end{figure}

Lastly, we explored the impacts of floe-eddy size ratios for the same collision count ranges.
While sea ice concentration can be maintained through various combinations of floe sizes and the number of floes, it represents only the combined effects of these factors and is insufficient as a standalone control parameter.
Instead, we used collision count as a control parameter to assess the influence of floe-eddy size ratios on ice floe rotation rates.
The PDFs of normalized rotation rates for different floe-eddy size ratios were compared in the collision count range $n_{col} = 70 \pm 10$ (figure \ref{Fig:TG_col_size}).
Overall, the distributions exhibit qualitatively similar trends to the free-drifting cases (figure \ref{Fig:TG_rot_PDF}).
The PDFs of floe rotation rates normalized by the averaged ocean vorticity exhibit distinct peak values and distributions for different floe-eddy size ratios (figure \ref{Fig:TG_col_size}$a$).
At $R_f/R_e=0.1$, the PDF peaks at unity but has a more dispersed distribution compared to $R_f/R_e=0.5$, where the PDF also peaks at unity. As $R_f/R_e$ increases to 1.0, the PDF peak shifts to 1.1.
The PDFs of floe rotation rates normalized by the center-of-mass ocean vorticity peak at lower rotation rates and show greater skewness for larger $R_f/R_e$ (figure \ref{Fig:TG_col_size}$b$).

\section{Conclusion} \label{sec:Conclusion}

We quantified the kinematic relationship between ocean eddies with surface expression and floes trapped within their cores and derived analytical relations linking their rotations and velocities, with floe-eddy size ratios as a key parameter in the analysis. Our results show that ice floes can act as vorticity meters of the ocean even though the combined effects of floe inertia and the filtering effect of ocean information over the floe area produce differences between direct measurements of ocean velocities and rotations and estimates derived from ice floe observations. These findings and the derived analytical relations demonstrate the potential applicability of our methodology for inferring ocean eddy characteristics from ice floe remote sensing measurements.

We began our analysis with a TG vortex, employing two distinct vorticity metrics: averaged ocean vorticity over the floe area and ocean vorticity at the floe center of mass.
Our analysis revealed that individual ice floes typically follow closed-loop trajectories, driven by the interplay of sea ice-ocean drag and a resultant force comprising the pressure gradient force due to sea surface tilt and the Coriolis force.
Along these trajectories, the rotation rates and velocities of ice floes show oscillatory behavior, undergoing alternating periods of acceleration and deceleration relative to the equilibrium values of force and torque.
The rotation rate normalized by the averaged ocean vorticity increases with increasing floe-eddy size ratio for both centered and off-centered ice floes. This trend arises from the gradual decrease in local ocean vorticity from the center of the TG vortex toward its boundaries. Ice floes behave as passive tracers when $R_f/R_e \leq 0.7$; however, as the floe-eddy size ratio increases to 1.4, the PDF peaks of rotation rates normalized by the averaged ocean vorticity shift toward higher values, reaching approximately 1.8.
Around $R_f/R_e \approx 0.5$, the distribution narrows and exhibits higher peaks, indicating that a solid body rotation approximation effectively describes ice floe motion within this range. Conversely, the rotation rate normalized by the center-of-mass ocean vorticity decreases for large floe-eddy size ratios. The velocities of ice floes were also used to estimate the averaged ocean vorticity within regions enclosed by their trajectories. 
These estimates were smaller than the true ocean vorticity of the same region; their ratios can be described by the relationship between ice floe and ocean velocities.

To further understand these dynamics, we derived analytical relations for ice floe rotation rates and velocities using the Taylor series expansion and the square-shape approximation. 
While the Taylor series expansion agrees well with the PDF peaks for small floe-eddy size ratios ($R_f/R_e \leq 0.5$), the square-shape approximation aligns closely with the PDF peaks across a broader range of size ratios.
Analytical relations for the ratio between the averaged ocean vorticity estimate and true ocean vorticity also show good agreement with the PDF peaks.
These relations were applied to a more realistic ocean eddy field obtained from a two-layer QG model. The trends observed in the TG vortex are also evident in the QG flow field when floe size is normalized by the trajectory-derived length scale. As the normalized floe size increases, the rotation rates normalized by the averaged ocean vorticity increase, while those normalized by the center-of-mass ocean vorticity decrease. The derived relations closely match the PDF peaks for normalized rotation rates, demonstrating their potential applicability in inferring ocean kinematics through ice floe motions.
 
We closed our study by exploring other factors influencing the rotational relationship between ice floes and the underlying ocean, such as ice floe thickness, atmospheric winds, and floe-floe collisions.
For ice floe thicknesses ranging from 0.1 to 1 m, we found that thicker floes exhibit reduced PDF peaks and more dispersed distributions due to increased floe inertia. 
Interestingly, the impact of varying floe thickness lessens with increasing floe-eddy size ratios and becomes negligible at $R_f/R_e=0.5$.
We also examined atmospheric wind forcing under low and high wind conditions. Under low wind conditions, wind-sea ice stresses have a minor effect on the normalized rotation rates, causing their PDF peaks to shift slightly to higher values. In high wind conditions, the PDFs of normalized rotation rates are skewed as stress ratios increase. The PDFs converge under larger stress ratios for increasing floe-eddy size ratios ($R_f/R_e=$0.1 to 1.0).
Finally, we investigated the impact of floe-floe collisions, which weaken the rotational link between ice floes and the underlying ocean.
For $R_f/R_e=0.5$, higher sea ice concentrations and larger collision counts lead to reduced PDF peaks and more dispersed distributions. When considering a fixed collision count ($n_{col}=70\pm10$), the PDFs for different floe-eddy size ratios exhibit trends consistent with those observed in free-drifting cases.

\subsection{Directions for future research}

Our idealized simulations effectively capture the key rotational relationship between ice floes and the underlying ocean eddies. However, incorporating additional properties of ocean and ice floes may enhance the estimation accuracy of our framework for practical applications in ocean characterization from ice floe remote sensing measurements.

The derived relations in the present study are applicable to both mesoscale and submesoscale motions in the ocean surface layer. These relations are particularly well-suited for analyzing eddies in the mesoscale regime ($Ro$ $\ll$ 1), where flows are predominantly horizontal and close to geostrophic balance \citep{Taylor2023}, resulting in rotation-dominant behaviors. In the submesoscale regime ($Ro$ $\sim$ 1), vertical motions and ageostrophic effects, such as convergence and divergence of flows, may introduce additional uncertainties into estimating ocean vorticity based on ice floe rotation rates. However, submesoscale motions still exhibit surface expressions at length scales comparable to the Rossby radius (5$–$15 km), similar to the smaller end of observed ice floe sizes (4$-$70 km) in Lagrangian ice floe datasets \citep{Lopez2019,Manucharyan2022b}, for which our framework is designed. At these scales, rotational flows remain more dominant than converging or diverging flows, preserving their two-dimensional characteristics. However, a comprehensive assessment of the applicability of the derived relations to submesoscale motions is needed to improve accuracy in this regime.

In the Arctic ocean, inertial oscillations and internal tidal motions, occurring roughly twice a day \citep{Gimbert2012,Watkins2023,Yuan2023}, may influence the motion of small-scale ice floes on sub-daily timescales. However, their effects on applying our analytical relation and simulation results to satellite observation datasets are negligible. Our framework is designed for datasets with daily resolution, capturing floes with a size down to about 5 km. At these temporal and spatial scales, inertial and tidal oscillations are expected to manifest as noise. Moreover, these oscillations primarily affect the trajectories of ice floes rather than their rotation rates about the center of mass, with minimal impact on the rotational relationship between ice floes and the underlying ocean. As part of future work, the effects of inertial and internal tidal oscillations will be further investigated using high-resolution ocean-sea ice models or in situ buoy measurements, which provide higher temporal and spatial resolutions.

In addition to ocean characteristics, ice floes indeed show variability in their physical properties, such as non-circular shapes (e.g., polygons) and rheology. During summer, large ice packs crack and fracture along preferential directions, generating diverse floe sizes and shapes, which in turn influence ice floe rotation rates. Ice floes with anisotropic shapes would exhibit distinct rotational behaviors, which can be expected from the motion of small, micro-scale particles in turbulent flows \citep{Parsa2012,Voth2017}. These particles undergo rapid orientation changes due to intermittently large velocity gradients and tend to align with these gradients in the flow. A similar preferential alignment is expected for ice floes, which may orient along large velocity gradients in the flow, thereby influencing their rotation rates. Further investigation is needed to fully understand this effect.

Furthermore, the rheological properties of ice floes can influence their rotational motion. Collisions between ice floes, as well as the convergence and divergence of underlying oceanic flows, can induce internal stress within floes, leading to deformation \citep{Gimbert2012,Yuan2023}. This deformation can create inhomogeneities in ice floe thickness and surface topography \citep{Feltham2008}, complicating the rotational relationship between ice floes and the ocean. Such effects may broaden the PDFs of the rotational relationship in our framework, increasing uncertainties in estimating ocean vorticity from ice floe rotation rates. However, in regions with low sea ice concentrations, where floe-floe collision rates are relatively low, treating ice floes as solid objects effectively captures the key features of the relationship. In addition, even in areas with high sea ice concentrations, our framework remains applicable to free-drifting ice floes--experiencing nearly zero internal stresses due to their consistent and unidirectional motion--for characterizing large-scale ocean currents, as well as local eddies.

While the present study focuses on the dynamical behavior of sea ice driven by ice-ocean drag, the thermodynamics of sea ice can influence the ice-ocean boundary layer, primarily through melting. Melting ice floes typically lower the surface temperature beneath them. During summer, substantial melting deposits freshwater to the ocean surface layer, creating sharp surface buoyancy gradients that drive mixed-layer instabilities \citep{Shrestha2022,Gupta2022}. These instabilities, in turn, generate submesoscale eddies and filaments near floe boundaries, which propagate from the ice edge toward floe centers. Such eddies can also transport ice floes into the open ocean, enhancing melting. In addition, induced vertical heat transport due to submesoscale motions may further increase ice floe melt rates \citep{Horvat2016,Gupta2022,Manucharyan2022a}. However, the effects of sea ice melting on the ice-ocean boundary layer become significant about a month after the onset of melting and have a reduced influence at low sea ice concentrations \citep{Gupta2022}; the conditions under which our framework is designed to operate. Furthermore, our framework focuses on ice floes during the spring-to-summer transition, whereas melting is generally more pronounced later in the summer.

Our findings and analytical relations provide a new framework for estimating local ocean vorticity with associated uncertainties. By incorporating Lagrangian observations of ice floe trajectories, rotation rates, velocities, and shapes, the analytical relations can provide locally averaged ocean vorticities and velocities. Furthermore, PDFs of ice floe rotation rates enable the assessment of uncertainties in estimated ocean information. 
Beyond kinematic analyses, ice floe rotation rates, which contain local, spatial ocean information, have the potential to provide insights into ocean kinetic energy and enstrophy spectra, thereby aiding in the characterization of energy cascades in mesoscale and submesoscale ranges. While the derived relations effectively describe the ice floe-ocean relationship, practical applications require further consideration of factors, including ocean characteristics (e.g., strong currents, eddy shapes, inertial and internal tidal oscillations, and mesoscale and submesoscale motions) and ice floe characteristics (e.g., shapes, thickness, surface topography, rheology, deformation, and thermodynamics).
Nonetheless, this study lays a foundation for a robust framework to characterize eddies from satellite remote sensing observations of sea ice motion, which can easily be extended beyond the Beaufort Gyre to include MIZ eddies in the Arctic and Antarctic Oceans.




\backsection[Acknowledgements]{M.K. sincerely thanks Dr. Daniel Watkins for his valuable discussions and insightful feedback.}

\backsection[Funding]{{M.K. and M.M.W were supported by the Office of Naval Research (ONR) Arctic Program (N00014-20-1-2753, N00014-22-1-2741, and N00014-22-1-2722), the ONR Young Investigator Program (N00014-24-1-2283), and the National Aeronautics and Space
Administration Science of Terra, Aqua, and Suomi-NPP Program (80NSSC22K0620). M.K., M.M.W, and G.E.M were also supported by the ONR Multidisciplinary University Research Initiatives Program (N00014-23-1-2014). This research used computational resources and services at the Center for Computation and Visualization, Brown University.}}

\backsection[Declaration of interests]{The authors report no conflict of interest.}


\backsection[Author ORCIDs]{M. Kim, https://orcid.org/0000-0002-8613-0206; G.E. Manucharyan, https://orcid.org/0000-0001-7959-2675; M.M. Wilhelmus, https://orcid.org/0000-0002-3980-2620.}


\appendix

\section{Identification of ice floes trapped by an ocean eddy} \label{sec:app_eddy_identification}

We leverage ice floe trajectories resembling closed loops to identify floes trapped in an eddy core, in agreement with conventional eddy detection algorithms \citep{Chelton2011,Mason2014}.
Several physical criteria were established for selecting closed-loop trajectories based on previous work using processed satellite remote sensing observations via the Ice Floe Tracker algorithm \citep{Lopez2019,Lopez2021}.
First, the curvatures of all daily segments along the trajectory must be of the same sign, indicating that an ice floe has been rotating clockwise or counterclockwise. Then, considering a nominal eddy size of 20 km, trajectory curvatures are evaluated as they should surpass a predefined threshold typically set at 0.05 [km$^{-1}$]. 
A third condition is set by the ratio of the arc length along the trajectory to the distance between the initial and final points. This value must exceed a specific threshold, typically set at 3, based on the geometry of a half-closed circular loop.
Finally, only trajectories with lifetimes greater than four simulation days are considered to ensure adequate data points. Only trajectories meeting all four criteria are classified as closed-loop trajectories.

\section{Effects of ice-ocean stress parametrization on the relationship between ocean vorticity and the rotation rate of ice floes}\label{sec:app_RK}

\begin{figure}
  \centerline{\includegraphics[width=30pc]{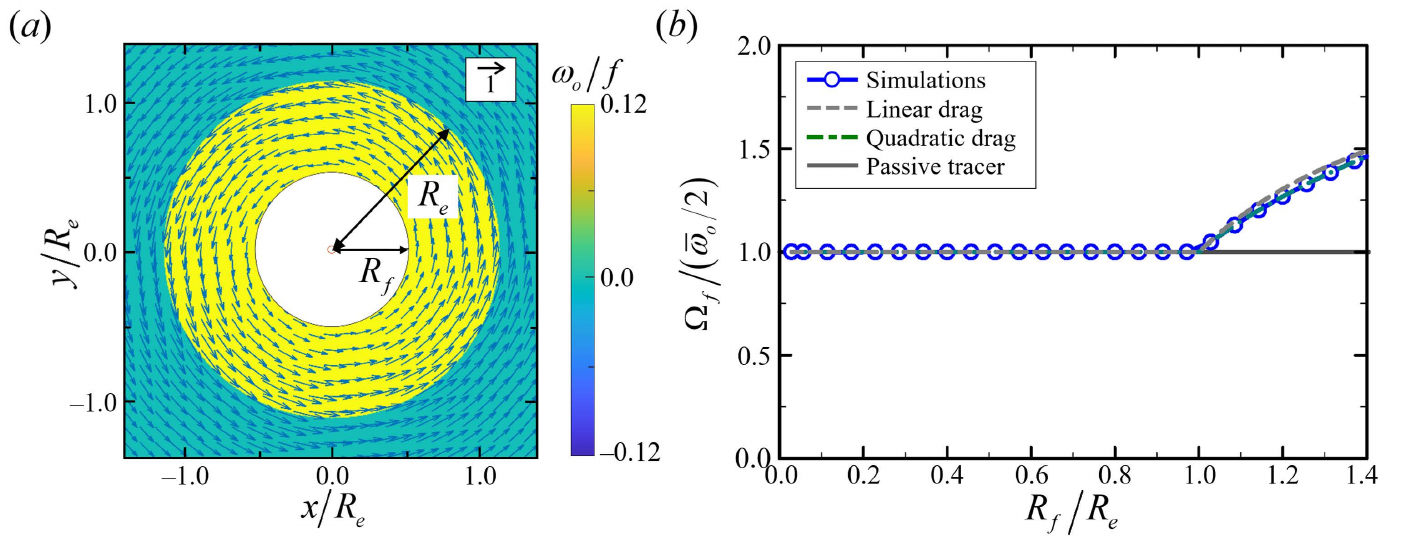}}
  \caption{Motions of ice floes in a Rankine vortex field. ($a$) An ice floe (white circle) positioned at the center of the vortex core (yellow region), with a floe-eddy size ratio of $R_f/R_e = 0.5$. The colors and arrows in the figure correspond to the magnitude of the vorticity normalized by the Coriolis parameter and the magnitude and direction of the velocity at a given location, respectively. ($b$) Rotation rates of the ice floe normalized by the averaged ocean vorticity over the floe area for different floe-eddy size ratios. The simulation results are compared with the analytical relations derived using the linear (dashed line) and quadratic (dashed-dotted line) drag laws, as well as with the passive tracer case (solid line).} \label{Fig:RK_center}
\end{figure}

The idealized Rankin (RK) vortex was used to examine the effect of ice-ocean stress on the rotational relationship between ice floes and the underlying ocean.
In the RK vortex, vorticity remains constant within the core region ($r \leq R_e$) and is zero in the outer region.
The velocity field is defined as $u_\theta = \Omega_f r$ for $r \leq R_e$ and $u_\theta = \Omega_f(R_e^2/r)$ for $r > R_e$, where $u_\theta$ is the azimuthal velocity. 
By representing the sea ice-ocean drag term using a linear drag law in equations \eqref{eq:sea-ice-eqns_torque} and \eqref{eq:sea-ice-eqns_torque_non}, we can derive the relation for ice floe rotation as follows:
\begin{equation} \label{eq:RK_rot_lin}
    \Omega_f = \frac{\overline{\omega}_o}{2} \quad \text{for } R_f \leq R_e, \qquad
    \Omega_f = \frac{\overline{\omega}_o}{2} \left[2 - \left(\frac{R_e}{R_f}\right)^2 \right] \quad \text{for } R_f > R_e,
\end{equation}
where $\overline{\omega}_o$ is the spatially averaged ocean vorticity over the floe area.
For $R_f \leq R_e$, $\overline{\omega}_o = 2\Omega_f$, whereas for $R_f > R_e$, $\overline{\omega}_o = 2\Omega_f(R_e/R_f)^2$. 
These relations suggest that when $R_f \leq R_e$, the rotation of the ice floe mirrors the average rotation of the underlying eddies, akin to the passive tracer case. However, when $R_f > R_e$, the rotation rate increases and tends to converge toward twice the averaged ocean rotation as $R_f$ approaches infinity. 
With the quadratic drag law (equation \ref{eq:stress_quad}), the rotation rate of ice floes remains unchanged compared to the linear parameterization, provided that $R_f \leq R_e$.
For $R_f > R_e$, the rotation rate of ice floes can be determined by solving the following algebraic equation:
\begin{equation}
    \Omega_f^{5/2} - \Omega_f^{3/2} \left(\frac{\overline{\omega}_o}{2}\right) \left[\frac{10}{3} + \frac{4}{3} \left(\frac{R_e}{R_f}\right)^3 \right] + \Omega_f^{1/2} \left(\frac{\overline{\omega}_o}{2}\right)^2 \left[4\left(\frac{R_e}{R_f}\right) + 5 \right] - \frac{16}{3} \left(\frac{\overline{\omega}_o}{2}\right)^{5/2} = 0,
\end{equation}
where $\Omega_f = \overline{\omega}_o/2$ for $R_f = R_e$.
The quadratic drag parameterization exhibits a closer fit to the simulation results compared to a linear drag (figure \ref{Fig:RK_center}).
However, the analytical relations demonstrate minimal dependence on drag parametrizations, as evidenced by the marginal error in the linear drag case.

\bibliographystyle{jfm}
\bibliography{jfm}

\end{document}